\documentclass[useAMS,usenatbib]{mn2e}
\usepackage{txfonts}
\usepackage{natbib}
\usepackage{aas_macros}
\usepackage[all]{xy}

\usepackage{graphicx}

\usepackage{hyperref}
\usepackage[usenames,dvipsnames]{color}
\definecolor{darkblue}{rgb}{0,0,.5}
\hypersetup{bookmarksopen,
			pdfstartview={FitH},
			colorlinks=true,
			breaklinks=true,
			linkcolor=darkblue,
			menucolor=darkblue,
			urlcolor=Red,
			citecolor=Green,
			linkcolor=blue}
\usepackage[all]{hypcap}

\usepackage{rotating}

\def\del#1{{}}

\newcommand{\dd}{\mathrm{d}}
\newcommand{\eqref}[1]{(\ref{#1})}

\newcommand{\n}{\hat{\bmath n}}

\newcommand{\dwl}{$3d$ weak lensing }
\newcommand{\sh}[1]{{}_{#1}Y_{\ell m}(\hat{\bmath{n}})}

\title[Intrinsic alignments in $3d$]{Intrinsic alignments and $3d$ weak gravitational lensing}
\author[Philipp M. Merkel and Bj{\"o}rn Malte Sch\"afer]
{Philipp M. Merkel$^1$\thanks{e-mail: philipp.merkel@urz.uni-heidelberg.de} and Bj{\"o}rn Malte Sch\"afer$^2$\\
${}^1$Institut f{\"u}r Theoretische Astrophysik, Zentrum f{\"u}r Astronomie, Universit{\"a}t Heidelberg, Philosophenweg 12, 69120 Heidelberg, Germany\\
${}^2$Astronomisches Recheninstitut, Zentrum f{\"u}r Astronomie, Universit{\"a}t Heidelberg, M{\"o}nchhofstra{\ss}e 12, 69120 Heidelberg, Germany}

\begin{document}
\pagerange{\pageref{firstpage}--\pageref{lastpage}}
\pubyear{2013}
\maketitle
\label{firstpage}

\begin{abstract}
 In this paper we show how intrinsic alignments can be incorporated consistently in the formalism of $3d$ weak lensing.
 We use two different descriptions of the intrinsic galaxy ellipticities, the so-called linear and quadratic model, respectively. 
 For both models we derive the covariance matrix of the intrinsic alignment signal (\textit{II}-alignments) and \textit{GI}-alignments (the cross correlation 
 of intrinsic and lensing induced ellipticities). 
 Evaluating the covariance matrices for the linear model numerically and comparing the results to the cosmic shear signal we find that for a low redshift 
 survey the total covariance matrix is dominated by the contributions of the \textit{II}-alignments.
 For a \textit{Euclid}-like survey \textit{II}-alignments still dominate over \textit{GI}-alignments but they are more than one order of magnitude smaller 
 than the lensing signal.
 The shape of the ellipticity covariance matrices is quite different in the $k$-$k'$-plane for cosmic shear on the one hand and intrinsic 
 alignments on the other hand. In comparison to lensing both alignment types tend to be rather elongated along the diagonal $k=k'$. Moreover, for high 
 multipoles ($\ell \sim 100$) intrinsic alignments are strongly concentrated along that diagonal.
\end{abstract}


\section{Introduction}
\label{sec_introduction}

Forthcoming weak lensing surveys like the \textit{Euclid} mission aim to constrain cosmological parameters to an accuracy of a few per cent.
Information from gravitational lensing \citep{1991MNRAS.251..600B,1994CQGra..11.2345S,1998MNRAS.301.1064K} is eminently important because it 
is complementary to that obtained from the cosmic microwave background \citep{2002PhRvD..65b3003H}. Being sensitive to all forms of matter and 
particularly to its time evolution, gravitational lensing experiments are essential in the investigation of dark energy; the most significant parameters are its 
equation of state along with its time evolution as well as the dark energy contribution to the Universe's matter content.

Conventionally, weak lensing or cosmic shear analyses are carried out in two dimensions. The main observables, angular correlation functions or 
equivalently angular power spectra, are obtained from a projection along the line-of-sight. Furthermore, one usually neglects the curvature of the sky 
considering only small patches which can be safely treated as flat. In order to overcome the line-of-sight projection and to make the valuable information 
contained in the time evolution of the lensing signal available, more advanced techniques have been developed in the past. 
These techniques take the photometric redshifts of the lensed galaxies into account. Simple stacking of usual (two-dimensional) lensing analysis carried 
out in several tomographic redshift slices already allows for tighter parameter constraints 
\citep{1999ApJ...522L..21H,2004ApJ...601L...1T,2006JCAP...06..025H}, especially in the case of the dark energy equation of state 
\citep{2002PhRvD..66h3515H}.
More elaborated than such tomographic methods is the three-dimensional formalism developed by \citet{2003MNRAS.343.1327H} and 
\citet{2005PhRvD..72b3516C}.
They completely account for the three-dimensional character of weak gravitational lensing by maintaining the radial dependence of the lensing potential 
throughout the whole analysis, i.e. by abandoning the necessity of any projection. Additionally, they provide access to the full sky. The cost of this very 
general formalism is its enhanced numerical complexity which is mainly due to the coupling of different density modes. This mode coupling  underlines 
the fact that the lensing potential is not statistically homogeneous.

For both, tomography as well as three-dimensional weak lensing, good knowledge of the lensed galaxies' redshift is indispensable. The error in 
photometric redshift estimation is therefore considered as one of the most important systematics in cosmic shear studies \citep{2008MNRAS.387..969A}. 
But the more the redshift accuracy improves, the more other systematical errors become significant.
Thus, the systematical error budget might soon be dominated by intrinsic shape correlations. The correlations in the shapes of galaxies, more precisely 
their ellipticities, are only an unbiased estimator of the weak gravitational shear field if the intrinsic shapes are completely randomly orientated. 
Otherwise, intrinsic shape correlations will mimic the lensing signal, thereby corrupting cosmological parameters derived from lensing studies which are 
carried out under the null hypothesis of absent intrinsic alignments 
\citep{2007NJPh....9..444B,2010A&A...523A...1J,2010MNRAS.408.1502K,2012arXiv1207.5939C}.
Employing shear tomography \citet{2007NJPh....9..444B} and \citet{2012MNRAS.424.1647K} showed that in particular the equation of state 
parameter of dark energy is severely biased. \citet{2008MNRAS.389..173K} obtained similar results using $3d$ cosmic shear.

Intrinsic shape correlations have been observed in a number of Sloan Digital Sky Survey samples 
\citep{2006MNRAS.367..611M,2011MNRAS.410..844M,2007MNRAS.381.1197H,2009ApJ...694..214O} as well as in numerical simulations 
\citep{2000MNRAS.319..649H,2006MNRAS.371..750H,2007ApJ...671.1135K}.

In order to remove, or at least reduce, the intrinsic alignment contamination of cosmic shear data a large variety of methods have been proposed and 
successfully applied. While \citet{2002A&A...396..411K,2003A&A...398...23K} and \citet{2003MNRAS.339..711H} exploit the locality of the intrinsic shape 
correlations to suppress the intrinsic alignment signal; the nulling techniques invented by \citet{2008A&A...488..829J,2009A&A...507..105J} modify the 
lensing efficiency in such a way that the resulting cosmic shear measures are free from intrinsic alignment contributions. 
Both approaches provide unbiased estimates of cosmological parameters at the expense of precision due to the unavoidable loss of information 
inherent to these methods. Alternatively, one can circumvent data rejection by explicitly including intrinsic alignments in the analysis. The additional 
parameters used for modelling the intrinsic alignments then need to either be determined from the data or marginalised over, thereby weakening the 
constraints on the derived cosmological parameters 
\citep{2005A&A...441...47K,2007NJPh....9..444B,2009ApJ...695..652B,2010A&A...523A...1J,2013MNRAS.tmp.1297H}.

All these methods, however, are tomographic methods. 
So far no comparable efforts have been made in case of $3d$ cosmic shear.
This is particularly deplorable because $3d$ weak lensing improves upon tomographic analyses as it avoids the loss of information going along with any 
binning in redshift.
Therefore, fully three-dimensional studies are extremely promising for measuring dark energy properties; very precise forecasts for the dark energy 
equation 
of state parameter and its time evolution can be achieved \citep{2006MNRAS.373..105H}.
Furthermore, $3d$ weak lensing is less sensitive to errors in the photometric redshifts of galaxies than tomography.
This makes intrinsic alignments very likely to dominate the systematical error budget in future $3d$ cosmic shear surveys.
Thus, reliable control of this systematic error is increasingly important.

Nonetheless the development of appropriate methods has received little attention in the past. This might be due to the fact that
a consistent three-dimensional description of intrinsic alignments derived from physical alignment models is not yet available.
Previous work analysing the impact of intrinsic alignments on $3d$ weak lensing measurements \citep[e.g.][]{2008MNRAS.389..173K} used the fitting 
formulae provided by \citet{2006MNRAS.371..750H}. 
These intrinsic alignment parametrisations are derived from numerical simulations.
In this paper we aim at a more analytical description. We establish the constitutive formalism necessary to incorporate intrinsic alignments into the 
framework of $3d$ weak lensing starting from physical alignment models. We concentrate on the two models proposed by 
\citet{2001MNRAS.320L...7C} and \citet{2004PhRvD..70f3526H} which have already been extensively used in the tomographic studies of intrinsic 
alignments mentioned above.
We present expressions for the resulting covariance matrices of the intrinsic alignment signal and for its cross correlation with that of cosmic shear. 
Comparison to the two-point statistics of weak cosmic shear then yields first insights into the strength of intrinsic alignment contamination, in particular its 
dependence on scale. Additionally, we investigate how this contamination varies with the depth of the  lensing survey under consideration.

We structured this paper as follows: In Section~\ref{sec_intrinsic_galaxy_ellipticity_and_lensing} we start with a brief recapitulation of the key quantities 
in the description of galaxy ellipticities and their relation to (weak) gravitational lensing. Subsequently, we present in detail the two intrinsic ellipticity 
models considered in this work. These models are then reformulated in the language of $3d$ cosmic shear in Section~\ref{sec_3d_formalism}. 
Furthermore, we derive expressions for the resulting covariance matrices of the various intrinsic alignment types. 
Section~\ref{sec_numerical_results} is devoted to the numerical evaluation of the covariance matrices derived before.
Focusing on the linear alignment model, we illustrate our results for two different lensing survey specifications.
Finally, we summarise our results in Section~\ref{sec_summary}.

Throughout this work we assume a spatially flat $\Lambda$CDM universe characterised by the \textit{WMAP7} best fit parameters 
\citep{2011ApJS..192...18K}. Thus, the contribution of matter to the Universe's energy budget is given by $\Omega_{\mathrm{m}}=0.3$, whereas the 
cosmological constant contributes with $\Omega_{\Lambda}=0.7$. The primordial perturbations in the cold dark matter (CDM) component with (scalar) 
fluctuation amplitude $\Delta^2_{\mathcal R}=2.43\times 10^{-9}$ (corresponding to $\sigma_8=0.8$) are nearly scale invariant ($n_\mathrm{S}=0.963$). 
Finally, the Hubble rate observed today is $H_0=100\, h\, \mathrm{km}\, \mathrm{s}^{-1}\,\mathrm{Mpc}^{-1}$ with $h=0.7$.

\section{Intrinsic galaxy ellipticity and gravitational lensing}
\label{sec_intrinsic_galaxy_ellipticity_and_lensing}

In projection, the shape of a galaxy on the sky can be characterized by its (complex) ellipticity 
\begin{equation}
 \epsilon = \epsilon_+ + \mathrm{i} \epsilon_\times = |\epsilon| \mathrm{e}^{2\mathrm{i} \varphi}.
\end{equation}
The real and imaginary part quantify the elongation and compression along two directions separated by $45^\circ$, respectively. The angle~$\varphi$ 
measures the misalignment between the local frame and the $x$-axis on the sky plane (assuming the line-of-sight along the $z$-axis).
Obviously, the ellipticity is a spin-2 field. It is invariant under rotations of the coordinate frame by an angle of $\upi$. Accordingly, its complex conjugate 
carries spin weight $-2$.
Under idealised conditions, i.e. in the absence of pixelisation, convolution and noise, the ellipticity components can be inferred from the second moments 
of the brightness distribution. The processing of real data, however, requires more involved methods and a variety of complementary shape 
measurement tools has been developed 
\citep{1995ApJ...449..460K,2003MNRAS.338...35R,2003MNRAS.338...48R,2007MNRAS.382..315M,2008MNRAS.390..149K,2011MNRAS.412.1552M}. 
A compilation of the different methods and their individual performance can be found in the handbooks of the GRavitational lEnsing Accuracy Testing 
Challenges 2008 and 2010 \citep{2009AnApS...3....6B,2010arXiv1009.0779K}.

Light rays reaching the observer from distant galaxies are deflected by gradients in the gravitational potential of the intervening large-scale structure. 
Consequently, the actually observed ellipticity of a galaxy $\epsilon^{(\mathrm o)}$ is made up of two contributions
\begin{equation}
 \label{eq_observed_ellipticity}
 \epsilon^{(\mathrm o)} = \epsilon + \gamma
\end{equation}
with
\begin{equation}
  \gamma = \gamma_+ + \mathrm{i} \gamma_\times.
\end{equation}
The shear $\gamma$ encodes the deformation of the source due to gravitational lensing 
\citep[see][for reviews]{2001PhR...340..291B,2008ARNPS..58...99H,2010CQGra..27w3001B}. 
By construction, it constitutes a spin-2 field, too.
In addition to the deformation, lensing also changes the size of the source isotropically. This effect is captured by the convergence $\kappa$ which is a 
real scalar quantity in contrast to the shear. The convergence is only directly observable in rare cases.

The simple superposition of intrinsic and lensing induced ellipticity in equation~\eqref{eq_observed_ellipticity} only holds in the limit of \emph{weak} 
gravitational lensing. Here the lensing effect on the individual galaxy is small, i.e. $|\gamma| \ll 1$ and $\kappa \ll 1$ and the reduced 
shear~$g\equiv\gamma/(1-\kappa)$ mediating between lensing and observed ellipticity
\begin{equation}
 \epsilon^{(\mathrm o)} = \frac{\epsilon+ g}{ 1 + g^*\epsilon}
\end{equation}
\citep{1997A&A...318..687S} can be well approximated by $\epsilon + \gamma$. Since the individual distortions are so small the observable weak 
lensing signal is obtained as a statistical average over an ensemble of galaxies. Equation~\eqref{eq_observed_ellipticity} underlines that the observed 
ellipticities are an unbiased estimator of the gravitational shear provided that the intrinsic shapes of the observed galaxies are randomly orientated and 
as long as lensing effects are weak, i.e. the approximation
\begin{equation}
 \epsilon^{(\mathrm o)} \sim \epsilon + g \sim \epsilon + \gamma
\end{equation}
holds.

The intrinsic galaxy shapes emerge, at least partially, from the cosmic tidal field introducing correlations in the intrinsic shape of nearby galaxies, 
so-called \textit{II}-alignments.
Since both tidal field and gravitational lensing have the same origin, namely the gravitational potential, there may also be correlations between the 
intrinsic galaxy shapes and those being lensing induced. These correlations are commonly referred to as \textit{GI}-alignments 
\citep{2004PhRvD..70f3526H}.
It is interesting to note that the two different alignment types feature opposed redshift dependencies. \textit{II}-alignments only arise from galaxy pairs 
which are physically close, i.e. their separation needs to be small both on the celestial sphere and in redshift, while the correlation length of 
\textit{GI}-alignments in redshift is compatible with that of weak lensing. This complicates the removal of \textit{GI}-alignments from cosmic shear data 
enormously. Another important difference between \textit{II}- and \textit{GI}-alignments arises from the fact that the latter can be negative indicating an 
anti-correlation between intrinsic and lensing induced ellipticity. While a lensed background galaxy appears preferentially tangentially aligned with the 
lens the tidal field of the very same lens aims to radially align a close-by foreground galaxy.

Basically, there are two models used to describe the intrinsic alignment of galaxies. They are linear and quadratic in the tidal shear tensor, respectively.
These models are rather simple and account neither for baryonic physics nor for the details of the formation process which are assumed to play a major 
role in building up the shape of an individual galaxy and for the orientation of its disk. Yet, they provide a very general and physically intuitive 
parametrisation which is largely based on the symmetry or rather spin properties of the ellipticity.
This is why they are a viable tool for an analytical treatment. Despite their simplicity, we expect these models to provide valuable information about the 
statistics of intrinsic alignments as for those the properties of a single galaxy should be of minor importance.
For a more extensive discussion of the range of applicability of these models we refer to \citet{2002MNRAS.332..788M,2004PhRvD..70f3526H}.

\subsection{Linear model}

The simplest model for intrinsic alignments is linear in the shear field, 
\[\epsilon_+ = -C_1 \left(\partial_x^2 - \partial_y^2 \right) \Phi_\mathcal{S},\]
\begin{equation}
 \label{eq_linear_model}
 \epsilon_\times = -2 C_1 \partial_x\partial_y \Phi_\mathcal{S},
\end{equation}
assuming that the galaxy shape is at least partially determined by the shape of the dark matter halo the galaxy resides in. 
The host halo forms from gravitational collapse in a tidal field and is therefore expected to be triaxial and aligned with the principal axes of the tidal field.
To physically motivate the ansatz in equation~\eqref{eq_linear_model} one can consider the extremely simple case of a spherical arrangement of test 
particles moving in a spatially slowly varying potential $\Phi$. A Taylor expansion of the potential about the origin then reveals that its gradient leads to a 
constant gravitational field shifting the test particles as a whole while the shape of the sphere is distorted by the quadratic term, i.e. the tidal field 
\citep{2001MNRAS.320L...7C}.

This ansatz is usually used to describe the shape of elliptical galaxies expecting that the baryonic matter, i.e. stars, fills up the gravitational well. The 
derivatives in equation~\eqref{eq_linear_model} are comoving derivatives and
the proportionality constant $C_1$ is a free parameter of the model which may be determined empirically by comparison with observations 
\citep[cf.][]{2001MNRAS.320L...7C,2004PhRvD..70f3526H}.
Since we are interested in the intrinsic ellipticity of galaxies we have to apply an appropriate filter function to the Newtonian potential in 
equation~\eqref{eq_linear_model}, which we indicate by the subscripted $\mathcal S$. The smoothing scale has to be chosen in such a way that it cuts 
off fluctuations corresponding to galaxy-sized objects. For instance one could impose a Gaussian filter function, i.e. 
$\Phi_\mathcal{S}(\bmath k) = \Phi(\bmath k) \mathrm{e}^{-(kR)^2/2}$, 
and adjust the filter scale $R$ to the mass scale of galaxies via the mean cosmic matter density $\bar\rho_\mathrm{m}$, i.e. by setting
$M=4/3\upi \bar\rho_\mathrm{m}R^3$. Typical values for galaxies are $M\sim10^{11} \ldots 10^{12} \, \mathrm{M}_\odot$.
Moreover, we assume that galaxy formation takes place during matter domination so that the linear Newtonian potential becomes independent of the 
scale factor.

\subsection{Quadratic model}

In contrast to ellipticals the shape of spiral galaxies is believed to be determined by their angular momentum. Since the disk forms perpendicular to the 
spin axis the intrinsic ellipticity may be described by the following ansatz \citep{2001MNRAS.320L...7C}
\[\epsilon_+ =  f(L, L_z) ( L_x^2 - L^2_y ),\]
\begin{equation}
 \label{eq_ansatz_ellipticities_angular_momentum}
 \epsilon_\times = 2f(L,L_z)L_x L_y
\end{equation}
where $L$ denotes the magnitude of the angular momentum. In the special case of an ideally thin disk, the function $f$ is given by 
$f(L,L_z) = 1/(L^2+L_z^2)$ so that 
\[\epsilon_+ = \frac{\hat L_x^2 - \hat L^2_y }{1 + \hat L^2_z},\]
\begin{equation}
 \label{eq_ellipticities_angular_momentum}
 \epsilon_\times = \frac{2\hat L_x\hat L_y}{1+ \hat L^2_z}
\end{equation}
with the angular momentum direction $\hat{\bmath L} \equiv \bmath L /L$.
The finite thickness of a realistic disk can be accounted for by simply rescaling the expressions for the ellipticity by a constant factor, which weakens the 
dependence on the inclination angle, i.e. $f(L,L_z) = \alpha/(L^2+L_z^2)$ with $\alpha \sim 0.75$, \citep{2001ApJ...559..552C}. It is interesting to note 
that the modulus of the ellipticity in equation~\eqref{eq_ellipticities_angular_momentum} is constant. Hence, the ellipticity components just measure the 
galaxy's orientation. 
For actual computations it is more convenient to simplify equation~\eqref{eq_ansatz_ellipticities_angular_momentum} by assuming 
$f(L,L_z)=C_2=\mathrm{const.}$ Since $f$ determines how the ellipticity scales with the angular momentum one expects 
$C_2 \sim \sqrt{\langle\epsilon^2\rangle / \langle L^2 \rangle}$ \citep{2002MNRAS.332..788M}. This comprises the effect of the finite disk thickness, as 
well, which weakens the correlation.

Under the assumption that the galaxy's angular momentum follows largely that of its host halo, one can compute the angular momentum in the 
framework of tidal torque theory 
(\citealp{1949MNRAS.109..365H,1955MNRAS.115....3S,1969ApJ...155..393P,1970Afz.....6..581D,1984ApJ...286...38W}; 
see \citealp{2009IJMPD..18..173S} for a recent review). 
Here the angular momentum is made up by the product of the halo's moment of inertia and the tidal field exerted by the neighbouring large-scale 
structure. 

The applicability of this mechanism is limited. Recent results from cosmological simulations indicate that tidal torque theory breaks down
at the turn around and virialisation of the halo \citep{2002MNRAS.332..339P}. \citet{2013ApJ...766L..15L} suggest that from this point on the angular 
momentum may emerge from a vortical rather than a shear flow. In this case, shape correlations could be related to correlations in the vorticity
of the cosmic flow field and hence be described by a single dynamical field in a way similar to the linear model presented before.
Extensions to tidal torque theory are also required in order to explain the hierarchical 
spin alignment of dark matter haloes in the cosmic web \citep{2013arXiv1303.1590A}. Finally, angular momentum acquisition is further complicated by 
the baryonic matter component \citep{2013ApJ...769...74S}.

Despite these limitations the framework of tidal torque theory is very well suited for an analytical description of most of the fundamental properties of the 
spin of dark matter haloes.
Both, the tensor of inertia and the shear field, can be related to the cosmic density field whose statistical properties are well-understood in linear theory. 
The actual computation of the resulting angular momentum correlation functions is quite involved due to the high dimensionality of the underlying 
probability distribution \citep{1988MNRAS.232..339H,2012MNRAS.421.2751S}. In the analysis of intrinsic alignments one therefore usually exploits the 
different correlation lengths of the tensor of inertia and the shear field, respectively. While the correlations of the latter are typically long ranged, 
correlations in the inertia tensor primarily arise from smaller scales. Thus, neglecting any correlations between tidal field and inertia tensor one can first 
average over all possible orientations of the inertia tensor and subsequently over the realizations of the tidal field. 
Setting $f(L,L_z)=C_2$ in equation~\eqref{eq_ansatz_ellipticities_angular_momentum} as discussed above one finds
\begin{equation}
 \epsilon_+ = C_2 \left[(\partial^2_x\Phi_{\mathcal S})^2 - (\partial_y^2\Phi_{\mathcal S})^2\right],
 \end{equation}
\begin{equation}
 \epsilon_\times = 2C_2  \left(\partial^2_x \Phi_{\mathcal S} - \partial_y^2 \Phi_{\mathcal S} \right) \partial_x\partial_y \Phi_{\mathcal S}.
\end{equation}
These relations were first derived by \citet{2001MNRAS.320L...7C} while our presentation above follows more closely the argumentation of \citet{2002MNRAS.332..788M}.
As before, the derivatives are comoving derivatives and the constant of proportionality $C_2$ is a free parameter of the model which needs to be determined from observations.

\section{$3d$ formalism}
\label{sec_3d_formalism}

The lensing effect results from gradients in the Newtonian gravitational potential. It can therefore be related to the cosmic density field by means of the 
Poisson equation. 
In the standard procedure of cosmic shear analysis the shear field enters as a projection along the line-of-sight 
$\gamma(\chi,\vartheta,\varphi) \rightarrow \gamma(\vartheta,\varphi)$, thus providing an integrated  measurement of the evolution of the cosmic 
density field. The projection, however, removes valuable information contained in its evolution as it inevitably mixes different spatial scales. 
This loss of information can be avoided, or at least reduced, by employing tomographic methods 
\citep{1999ApJ...522L..21H,2002PhRvD..66h3515H,2004MNRAS.348..897T,2011MNRAS.413.2923K,2012MNRAS.423.3445S}. 
After dividing the population of lensed galaxies into redshift bins the projection can be carried out piecewise over smaller redshift intervals thereby 
increasing the sensitivity of the lensing signal.

The focus of this work rests on another possibility to overcome the limitations of the line-of-sight projection: $3d$ weak lensing 
\citep{2003MNRAS.343.1327H,2005PhRvD..72b3516C}.
This formalism allows for a direct three-dimensional mapping of the cosmic density field.

As detailed in the preceding section, in the two simple models we are considering in this work intrinsic galaxy alignments may also be related to the 
Newtonian gravitational potential. The linear model is very similar to the description of weak cosmic shear. We shall therefore start with a short repetition 
of the formalism of $3d$ weak lensing in a very general way allowing for a convenient extension to intrinsic alignments later on.

\subsection{Cosmic shear}

\dwl  is formulated in harmonic space. Since all fields of interest carry spin-weight 2, the most natural choice of basis functions is given by a combination 
of spherical Bessel functions $j_\ell(z)$ \citep{1972hmfw.book.....A} and spin-weighted spherical harmonics 
${}_{s}Y_{\ell m} (\bmath{\hat{n}})$ \citep{2000PhRvD..62d3007H}. Any spin-$s$ quantity can then be expressed as
\begin{equation}
 S(\bmath x) = \sqrt{\frac{2}{\upi}} \sum_{\ell m} \int k^2\dd k \, S_{\ell m} (k) {}_{s}Y_{\ell m} (\bmath{\hat{n}}) j_\ell(kx).
\end{equation}
Inverting the last equation yields the expansion coefficients
\begin{equation}
 S_{\ell m} (k ) = \sqrt{\frac{2}{\upi}} \int \dd^3 x \, S(\bmath x) j_\ell(kx) {}_{s}Y^*_{\ell m} (\bmath{\hat{n}}).
\end{equation}
Here and in the remainder of this work we will assume a spatially flat universe. The generalisation to a universe with arbitrary spatial curvature is 
straightforward. Most recently this spherical Fourier Bessel expansion has been applied in the analysis of baryon acoustic oscillations 
\citep{2013arXiv1301.3673P}.

The cosmic shear field, as well as the convergence, are derived from the lensing potential 
\begin{equation}
 \label{eq_lensing_potential}
 \phi(\bmath x) = \phi\left(\chi,\vartheta,\varphi\right)  = 2\int_0^\chi \dd \chi' \, \frac{\chi - \chi'}{\chi\chi'} \Phi \left(\chi',\vartheta,\varphi\right)
\end{equation}
by applying twice the eth operator $\eth$ \citep{1966JMP.....7..863N,1967JMP.....8.2155G} and its complex conjugate $\bar\eth$
\begin{equation}
 \label{eq_gamma_second_eth_derivative}
 \gamma (\bmath x ) = \frac{1}{2}\eth\eth\, \phi(\bmath x),
\end{equation}
\begin{equation}
  \kappa (\bmath x) = \frac{1}{4}\left(\eth\bar\eth + \bar\eth\eth \right) \phi(\bmath x) = \frac{1}{2}\Delta_{\vartheta\varphi}\phi(\bmath x)
\end{equation}
\citep{2005PhRvD..72b3516C}, respectively. 
The distances $\chi$ in equation~\eqref{eq_lensing_potential} are comoving distances as we assume a spatially flat universe. 

Applied to a spin-weighted spherical harmonic the eth operator acts as a spin raising operator
\begin{equation}
 \eth\, \sh{s} = \sqrt{(\ell - s ) (\ell + s + 1)}\, \sh{s+1},
\end{equation}
whereas its complex conjugate lowers the corresponding spin weight
\begin{equation}
 \bar\eth\, \sh{s} = -\sqrt{(\ell + s ) (\ell - s + 1)}\, \sh{s-1}.
\end{equation}
Noting that the lensing potential is a true scalar, i.e. spin-0 quantity, the following relations hold
\begin{equation}
 \label{eq_gamma_phi_relation}
 \gamma_{\ell m}(k) = \frac{1}{2} \sqrt{\frac{(\ell+2)!}{(\ell-2)!}}\phi_{\ell m}(k),
\end{equation}
\begin{equation}
\kappa_{\ell m}(k) = -\frac{\ell(\ell+1)}{2}\phi_{\ell m}(k).
\end{equation}
The coefficients of the lensing potential can be written in compact form as
\begin{equation}
 \label{eq_phi_Phi_relation}
 \phi_{\ell m}(k) = \eta_\ell(k,k') \Phi_{\ell m}^0 (k')
\end{equation}
by introducing the matrix
\begin{equation}
 \eta_\ell (k, k') \equiv \frac{4}{\upi} \int \chi^2 \dd \chi j_\ell(k\chi) \int_0^\chi \dd \chi' \frac{\chi-\chi'}{\chi\chi'} j_\ell(k'\chi') \frac{D_+(a')}{a'}
\end{equation}
along with the following summation convention
\begin{equation}
 \mathcal A(k,k') \mathcal B(k', k'') \equiv \int k'^2 \dd k'\, \mathcal A(k,k')\mathcal B(k',k'').
\end{equation}
The inclusion of the linear growth function $D_+(a)$ allows for the use of today's gravitational potential $\Phi^0_{\ell m}(k)$, which in turn is related to the density contrast via the (comoving) Poisson equation
\begin{equation}
 \Phi_{\ell m}(k) = -\frac{3\Omega_{\mathrm{m}}}{2\chi_H^2} \frac{1}{a} \frac{\delta_{\ell m}(k)}{k^2}.
\end{equation}
Here we have introduced the Hubble distance $\chi_H \equiv c/H_0$.

Having  established a relation between the harmonic transform of the gravitational shear and that of the cosmic density field (cf. 
equations~\ref{eq_gamma_phi_relation} and~\ref{eq_phi_Phi_relation}), we are still leaking a relation to the actually accessible observables in weak 
lensing measurements. Therefore, we consider the following discrete estimator
\begin{equation}
 \label{eq_estimator}
 \hat\gamma_{\ell m} (k) = \sqrt{\frac{2}{\upi}}\sum_{\mathrm{galaxies} \ g} \gamma_{\mathrm g}(\bmath x_{\mathrm g}) j_\ell (k\chi_g)\, 
 {}_2Y^*_{\ell m} (\bmath{\hat n}_\mathrm g)
\end{equation}
\citep{2003MNRAS.343.1327H} constructed from the gravitational shears of individual galaxies. In order to calculate the expectation value of this 
estimator, we have to take observational uncertainties, as well as some specifications of the galaxy survey, into account.

In any realistic experiment the exact radial position of a galaxy is not known, however an estimate of its photometric redshift may be available. The 
conditional probability for the estimated redshift $z$ given the true redshift $z_{\mathrm{true}}$ is typically modelled by a Gaussian
\begin{equation}
 p(\chi | \chi_{\mathrm{true}}) \dd\chi= \frac{1}{\sqrt{2\upi}\sigma_z}\exp\left[-\frac{(z - z_{\mathrm{true}})^2}{2\sigma_z^2}\right] \dd z.
\end{equation}
For simplicity, we assume rather optimistically $\sigma_z = 0.02$ independent of redshift.

The survey itself is assumed to be a full-sky experiment, i.e. it is homogeneous and isotropic in its angular part. The radial galaxy number density is 
described by
\begin{equation}
 n(z)\dd z \propto z^2 \exp\left[-\left( \frac{z}{z_0}\right)^\beta \right]\dd z.
\end{equation}
Having \textit{Euclid} in mind we expect 100 galaxies per square arcminute and set $z_0=0.64$ and $\beta = 3/2$ \citep{2012arXiv1206.1225A}.

After the transition to a continuos formulation the expectation value of the estimator is given by
\begin{equation}
 \label{eq_expectation_value_estimator_of_gamma}
 \bar{\gamma}_{\ell m}(k) = \mathcal Z_\ell ( k, k') \mathcal M_\ell ( k', k'') \gamma _{\ell m} (k''),
\end{equation}
where we have defined the two new matrices by
\begin{equation}
 \mathcal Z_\ell(k,k') \equiv \frac{2}{\upi} \int \chi'^2\dd \chi' \,j_\ell (k\chi') \int\dd\chi \, p(\chi'|\chi) j_\ell( k' \chi) 
\end{equation}
and
\begin{equation}
 \mathcal M_\ell ( k, k') \equiv \frac{2}{\upi} \int \chi^2 \dd \chi \, j_\ell ( k\chi) j_\ell(k'\chi) n(\chi).
\end{equation}
Being proportional to the linear density contrast the expectation value itself is obviously zero. Valuable information, however, can be extracted from its 
covariance. It reads in terms of the linear matter power spectrum $P(k)$
\begin{eqnarray}
 \nonumber
  C_\ell^{\gamma\gamma}(k, k') &=& \left\langle \bar\gamma_{\ell m}(k) \bar\gamma^*_{\ell' m'}(k') \right\rangle \\
 &=& A^2 \frac{(\ell +2)!}{(\ell-2)!}\mathcal B_\ell (k,k'')\frac{P(k'')}{k''^4} \mathcal B_\ell ( k', k'')\label{eq:cosmic_shear_cov_matrix}
\end{eqnarray}
with the constant $A\equiv 3\Omega_{\mathrm m} / 4\chi_H^2$ and the matrix 
$\mathcal B_\ell(k,k') \equiv \mathcal Z_\ell ( k, k'') \mathcal M_\ell( k'', k''') \eta_\ell ( k''',k')$.

In contrast to conventional cosmic shear studies which are confined to the two-dimensional sky, the covariance matrix of the three-dimensional lensing 
field acquires off-diagonal elements in $k$-space.
Accordingly, the shear field has different statistical properties with respect to the underlying cosmic density field. The latter is statistically homogeneous 
and isotropic characterised by a power spectrum which is diagonal in harmonic space, i.e.
\begin{equation}
 \left\langle \delta_{\ell m} (k) \delta^*_{\ell' m'} (k')\right\rangle = \frac{P(k)}{k^2}\delta_\mathrm{D} (k-k') \delta_{\ell\ell'}\delta_{m m'}.
\end{equation}
The lensing potential, as a projection along the line-of-sight is no longer statistically homogeneous and isotropic in radial direction. Conversely, the 
covariance matrix is still diagonal in multipole space highlighting that it does maintain statistical homogeneity and isotropy in angular direction at fixed 
radial position.

The structure of the covariance matrix becomes more involved if nonlinear structure growth is considered.
Beyond linear theory the growth of cosmic structures becomes scale dependent. Its time, or equivalently redshift, dependence cannot longer
be expressed in terms of the growth function. Accordingly equation~\eqref{eq_phi_Phi_relation} needs to be replaced by
\begin{equation}
 \phi_{\ell m}(k) = \frac{4}{\upi} \int \chi^2 \dd \chi j_\ell(k\chi) \int_0^\chi \dd \chi' \frac{\chi-\chi'}{\chi\chi'} j_\ell(k'\chi') \Phi_{\ell m}(k', \chi').
\end{equation}
The computation of the cosmic shear covariance matrix then requires the evaluation of the matter power spectrum for modes located at different 
(comoving) distances (or instances of time), i.e. a term of the form $P(k,\chi,\chi')$ enters equation~\eqref{eq:cosmic_shear_cov_matrix} inside the 
$\chi$-integrations.
For further progress \citet{2005PhRvD..72b3516C} suggested and justified to set $P(k,\chi,\chi') \simeq \sqrt{P(k,\chi) P(k,\chi')}$. From this it is easy to 
recover our expressions which are restricted to linear structure formation 
$P(k,\chi,\chi') = \sqrt{P(k,\chi) P(k,\chi')} = \sqrt{D^2_+(a)P(k) D^2_+(a')P(k)} = D_+(a) D_+(a')P(k)$. Thus, the power spectrum does not enter the
$\chi$-integrations.

Large-scale modes of the estimator~\eqref{eq_expectation_value_estimator_of_gamma} can be safely treated as Gaussian random fields, hence the 
corresponding likelihood is completely determined by the covariance matrix~\eqref{eq:cosmic_shear_cov_matrix}. Extending the Gaussian likelihood to 
small scale-modes, however, is expected to introduce a significant bias in the derived cosmological parameters
\citep{2009MNRAS.395.2065T,2013arXiv1301.3588S}.
While this is a consequence of nonlinear structure growth,
a non-Gaussian likelihood may also arise from the finite volume of any realistic survey. Here, large-scale modes, which exceed the survey, couple 
adjacent large wavelength perturbations. This results in an (additional) non-Gaussian signature on scales which are much smaller than the survey 
\citep{2013arXiv1302.6994T}.

\subsection{Intrinsic alignments}

For a realistic experiment the gravitational shear needs to be replaced by the observed ellipticity $\epsilon^{(\mathrm{o})}$ in the estimator given by 
equation~\eqref{eq_estimator}. In the absence of intrinsic alignments the covariance matrix of the gravitational shear field is almost unaffected by this 
substitution. It just acquires an additional shot noise term given by
\begin{equation}
 \mathcal{N}_\ell (k,k') = \frac{\sigma_\epsilon^2}{4} \mathcal{M}_\ell (k,k'),
\end{equation}
where $\sigma_\epsilon^2$ denotes the variance of the observed ellipticity, typically $\sigma_\epsilon^2\lesssim 0.1$ \citep{2003MNRAS.343.1327H}.

The situation, however, is completely different when the intrinsic galaxy shapes are correlated. Then, there are basically three distinct contributions to 
the total covariance (omitting the shape noise for simplicity in the following)
\begin{eqnarray}
 \nonumber
 \left\langle \epsilon^{(\mathrm o)}_{\ell m} (k) \epsilon^{(\mathrm o)*}_{\ell m} (k') \right\rangle
 &=&
 C_\ell(k,k')\\
 &=&
 C_\ell^{\gamma\gamma} ( k, k') + C_\ell^{\epsilon\epsilon} ( k, k') + C_\ell^{\epsilon\gamma} ( k, k').
\end{eqnarray}
Assuming a completely covered sky guarantees that the covariance matrix is diagonal in $\ell$ and $m$. The second term describes 
\textit{II}-alignments whereas the third one captures \textit{GI}-alignments. Actually, the last term solely arises from the correlation between the intrinsic 
shape of a foreground galaxy and the lensing induced shear of a background galaxy. Lensed galaxies are typically located far enough from the lens in 
order to safely consider intrinsic shape and tidal field being uncorrelated, i.e. their distance exceeds several correlation lengths.
Concentrating on linear theory, this term is only present in the linear model because in case of the quadratic model the covariance involves the 
bispectrum of the gravitational potential which identically vanishes for a Gaussian random field. In order to compute the individual contributions to the 
covariance we have to find the corresponding $3d$ expressions for the intrinsic ellipticities first.

\subsubsection{Linear model}

We start with the linear model as given in equation~\eqref{eq_linear_model}.
In a first step we convert the comoving derivatives into angular derivatives, i.e.
\begin{equation}
 \epsilon_+ = -C_1 \left(\partial_x^2 - \partial_y^2 \right) \Phi_{\mathcal S}
 = -\frac{C_1}{\chi^2}\left(\partial_{\theta_1}^2 - \partial_{\theta_2}^2 \right)\Phi_{\mathcal S}(\chi\btheta,\chi),
\end{equation}
\begin{equation}
 \epsilon_\times = -2 C_1 \partial_x\partial_y \Phi_{\mathcal S}
 = -2 \frac{C_1}{\chi^2} \partial_{\theta_1}\partial_{\theta_2} \Phi_{\mathcal S}(\chi\btheta,\chi),
\end{equation}
where $\btheta=(\theta_1,\theta_2)$ denotes a two-dimensional angular position on the sky.
In order to facilitate the incorporation of the intrinsic ellipticity field in the $3d$ formalism we then replace the angular derivatives by covariant derivatives 
on the (unit) sphere. This step might seem worrisome because the tidal shearing responsible for the intrinsic ellipticity is not confined to the surface of 
the observed sky. However, in practice we are only interested in patches of the sky small enough to be considered as flat, so that then covariant 
derivatives reduce to partial derivatives again. We denote a covariant derivative acting on a vector field by 
$\nabla_\mu X_\nu = \partial_\mu X_\nu - \Gamma^\alpha_{\mu\nu} X_\alpha$ (keeping Einstein's summation convention in mind) and recall that the 
non-vanishing Christoffel symbols on the two-dimensional sphere are $\Gamma_{\varphi\varphi}^\vartheta = -\sin\vartheta\cos\vartheta$ and
$\Gamma_{\vartheta\varphi}^\varphi = \Gamma_{\varphi\vartheta}^\varphi = \cot\vartheta$.
We then have
\begin{equation}
 \epsilon_+ = -\frac{C_1}{2\chi^2} \left(\nabla_\vartheta\nabla_\vartheta - \csc^2\vartheta\,\nabla_\varphi\nabla_\varphi \right)\Phi_{\mathcal S},
\end{equation}
\begin{equation}
 \epsilon_\times = -\csc\vartheta\, \frac{C_1}{\chi^2}\nabla_\vartheta\nabla_\varphi \Phi_{\mathcal S}.
\end{equation}
The additional $\csc\vartheta$ term in the last expression mirrors the fact that the two ellipticity modes form the components of a rank two tensor rather 
than those of a vector, i.e.
\begin{equation}
 \label{eq_ellipticity_tensor}
 \epsilon_{\mu\nu} = \left(
 \begin{array}{cc}
 \epsilon_+ & \sin\vartheta \, \epsilon_\times\\
 \sin\vartheta\,  \epsilon_\times & -\sin^2\vartheta \, \epsilon_+
 \end{array}
 \right),
\end{equation}
and that the basis vectors are orthogonal but not orthonormal.

The great advantage of reformulating the intrinsic ellipticity in terms of covariant derivatives is that we can now exploit its relation to the eth operator 
\citep{2005PhRvD..72b3516C}. It allows to write the (complex) galaxy's ellipticity in the most compact form, namely as second eth-derivative of a suitably 
constructed scalar potential
\begin{equation}
 \epsilon(\bmath x) = -\frac{C_1}{2}\eth\eth\, \Psi(\bmath x)
\end{equation}
with
\begin{equation}
 \label{eq_definition_of_psi}
\Psi(\chi,\vartheta,\varphi) \equiv \frac{\Phi_{\mathcal S} (\chi,\vartheta,\varphi)}{\chi^2}
\end{equation}
in complete analogy to the weak lensing case (cf. equation~\ref{eq_gamma_second_eth_derivative}), replacing the lensing potential by the auxiliary 
potential~$\Psi$. There is one important difference between the two potentials. In contrast to the lensing potential, the field~$\Psi$ does not evolve in 
time because we assume that galaxy formation takes place during matter domination. Accordingly, we find the following relation for the harmonic 
coefficients
\begin{equation}
 \epsilon_{\ell m} (k)= -\frac{1}{2} \sqrt{\frac{(\ell+2)!}{(\ell - 2)!}} \Psi_{\ell m} (k)
\end{equation}
which are connected to the smoothed Newtonian gravitational potential by
\begin{equation}
 \Psi_{\ell m} (k) = C_1 \tilde\eta_\ell(k,k')  \Phi^{\mathcal S,0}_{\ell m} (k')
\end{equation}
with
\begin{equation}
 \tilde{\eta}_\ell(k,k') = \frac{2}{\upi} \int \dd \chi j_\ell ( k\chi) j_\ell ( k'\chi).
\end{equation}
Due to the missing time dependence of the matrix $\tilde\eta_\ell(k,k')$ the matrix $\tilde{\mathcal{B}}_\ell(k,k')$, which we define in complete analogy to 
the cosmic shear case, takes the particularly simple form
\begin{equation}
 \tilde{\mathcal{B}}_\ell ( k, k') = \frac{2}{\upi} \int \dd \chi j_\ell( k \chi) \int \dd \chi' p(\chi| \chi') j_\ell (k' \chi') \frac{n (\chi')}{\chi'^2}.
\end{equation}
Finally, we arrive at the expressions for the covariance matrix of the \textit{II}-alignments
\begin{eqnarray}
 \nonumber
  C_\ell^{\epsilon\epsilon}(k, k') &=& \left\langle \bar\epsilon_{\ell m}(k) \bar\epsilon^*_{\ell' m'}(k') \right\rangle \\
 &=& A^2 \frac{(\ell +2)!}{(\ell-2)!}\tilde{\mathcal B}_\ell (k,k'')\frac{P(k'')}{k''^4} \tilde{\mathcal B}_\ell ( k', k'')
\end{eqnarray}
and of the \textit{GI}-alignments
\begin{eqnarray}
 \nonumber
  C_\ell^{\epsilon\gamma}(k, k') &=& \left\langle \bar\epsilon_{\ell m}(k) \bar\gamma^*_{\ell' m'}(k') \right\rangle \\
 &=& -A^2 \frac{(\ell +2)!}{(\ell-2)!}\tilde{\mathcal B}_\ell (k,k'')\frac{P(k'')}{k''^4} \mathcal B_\ell ( k', k'').
\end{eqnarray}
It is interesting to note that without the inverse $\chi^2$-weighting of the gravitational potential in equation~\eqref{eq_definition_of_psi} the 
matrix~$\tilde{\eta}$ would reduce to the unit matrix because of the orthogonality relation of the spherical Bessel functions 
(cf. equation~\ref{eq_orthogonality_relation_Bessel_functions}). In this case the intrinsic ellipticity field would be a statistically homogeneous field and the 
non-diagonal structure of  its covariance matrix would solely result from the construction of the estimator $\bar{\epsilon}_{\ell m}(k)$.

\subsubsection{Quadratic model}
\label{subsec_quadratic_model}

In order to find the $3d$ generalisation of the quadratic alignment model we could in principle just repeat the steps we have carried out for the linear 
model in the previous section. However, we can shorten the derivation by noting that
\begin{equation}
 \epsilon = \frac{C_2}{\chi^4} \left(\partial^2_{\theta_1} \Phi_{\mathcal S} + \partial^2_{\theta_2} \Phi_{\mathcal S}
 						\right)
						\left(\partial^2_{\theta_1} \Phi_{\mathcal S} - \partial^2_{\theta_2} \Phi_{\mathcal S} + 2 \mathrm{i} \partial_{\theta_1}
						\partial_{\theta_2}\Phi_{\mathcal S}\right).
\end{equation}
We then find
\begin{equation}
 \label{eq_epsilon_quadratic_model_eth}
 \epsilon(\bmath x) = \frac{C_2}{2} \bigl[\eth\eth\, \Psi (\bmath x)\bigr] \left[\left( \eth \bar{\eth} + \bar{\eth}\eth\right) \Psi (\bmath x)\right],
\end{equation}
where the potential $\Psi$ is defined as before (cf. equationº \ref{eq_definition_of_psi}).
Obviously, the ellipticity in equation~\eqref{eq_epsilon_quadratic_model_eth} has spin weight two. It is the most general quadratic function
of the tidal shear tensor with spin-2.

Aiming at the statistics of the intrinsic alignments we need to relate the harmonic coefficients of the ellipticity field to that of the gravitational potential 
next. This relation is much more involved than in case of the linear model. We start with the intermediate result
\begin{eqnarray}
 \nonumber
 \epsilon_{\ell m} ( k ) &=& (-1)^{m+1}\sqrt{\frac{2}{\upi^3}}C_2 \sum_{\ell_1\ell_2 m_1 m_2} \ell_1(\ell_1+1) \sqrt{\frac{(\ell_2+2)!}{(\ell_2-2)!}}\\ 
 \nonumber
	&& 	\times \, \mathcal{I}_{\ell_1\ell_2\ell} (k_1,k_2,k) \, {}_2\mathcal{W}^{\ell_1\ell_2\ell}_{m_1 m_2 m}\\
 \label{eq_harmonic_coeff_quadratic_model}
	&&	\times \, \tilde\eta_{\ell_1} ( k_1, k_3) \Phi^{\mathcal S,0}_{\ell_1 m_1}(k_3)
			\tilde\eta_{\ell_2} ( k_2, k_4) \Phi^{\mathcal S,0}_{\ell_2 m_2}(k_4).
\end{eqnarray}
Here we have introduced several new quantities which emerge from the quadratic structure in the potential. The mode coupling is captured by
\begin{equation}
 \mathcal{I}_{\ell_1\ell_2\ell} (k_1,k_2,k) \equiv
 \int \chi^2 \dd \chi\, j_{\ell_1} (k_1\chi) j_{\ell_2} (k_2\chi) j_\ell (k\chi)
\end{equation}
while the angular mixing matrix is given by
\begin{equation}
 \label{eq_angular_mixing_matrix}
 {}_2\mathcal{W}^{\ell_1\ell_2\ell}_{m_1 m_2 m} \equiv \Pi_{\ell_1\ell_2\ell}
 \left(
 \begin{array}{ccc}
 	\ell_1 & \ell_2 & \ell \\
	0 & -2 & 2
 \end{array}
 \right)
  \left(
 \begin{array}{ccc}
 	\ell_1 & \ell_2 & \ell \\
	m_1 & m_2 & -m
 \end{array}
 \right)
\end{equation}
with
\begin{equation}
 \Pi_{\ell_1\ell_2\ell}\equiv\sqrt{\frac{(2\ell_1+1)(2\ell_2+1)(2\ell+1)}{4\upi}}.
\end{equation}
The six-fold indexed quantities denote Wigner $3j$-symbols which are related to spherical harmonics by
\begin{eqnarray}
 \nonumber
\int\dd\Omega \; {}_{s_1}Y_{\ell_1m_1}(\n) \, {}_{s_2}Y_{\ell_2m_2}(\n)\,  {}_{s_3}Y_{\ell_3m_3}(\n) \qquad\qquad\qquad\qquad&&  \\
 \qquad = \Pi_{\ell_1\ell_2\ell_3}
 \left(
 \begin{array}{ccc}
 	\ell_1 & \ell_2 & \ell_3 \\
	-s_1 & -s_2 & -s_3
 \end{array}
 \right)
 \left(
 \begin{array}{ccc}
 	\ell_1 & \ell_2 & \ell_3 \\
	m_1 & m_2 & m_3
 \end{array}
 \right)&&
\end{eqnarray}
\citep{2000PhRvD..62d3007H}.
The corresponding expectation value of the estimator is then obtained from equation~\eqref{eq_expectation_value_estimator_of_gamma} 
replacing $\gamma_{\ell m}(k)$ by $\epsilon_{\ell m}(k)$. 

For the calculation of its covariance we need to evaluate the trispectrum of the Newtonian potential.
Since we are focusing on linear theory, the trispectrum is completely characterised by its reducible part and we can therefore invoke Wick's theorem. 
Here we just quote the final result relegating a detailed derivation to Appendix~\ref{sec_appendix}:
\begin{eqnarray}
 \label{eq_covariance_matrix_quadratic_model}
 \nonumber
 C^{\epsilon\epsilon}_\ell(k,k') &=& \frac{A^4C^2_2}{2\ell +1} \sum_{\ell_1\ell_2} \mathcal B^k_{\ell\ell_1\ell_2}(k_3,k_4) 
 \frac{P(k_3)}{k_3^4} \frac{P(k_4)}{k_4^4}\\
 &&
 \times \left(\mathcal{B}^{k'}_{\ell\ell_1\ell_2} (k_3,k_4 ) + (-1)^{\ell_1+\ell_2+\ell} \, \mathcal{B}^{k'}_{\ell\ell_2\ell_1}(k_3,k_4) \right).
\end{eqnarray}
The definition of the matrix $\mathcal B^k_{\ell\ell_1\ell_2}(k_3,k_4)$ is given in equation~\eqref{eq_def_B_matrix_quadratic_model}.
Note that the covariance matrix is diagonal in multipole space reflecting the fact that we have assumed full sky coverage.
While it is independent of $m_{1,2}$ the quadratic form of the intrinsic ellipticities requires the evaluation of the double sum in $\ell_{1,2}$ leaving the 
numerical effort for its computation tremendous (cf. Section~\ref{subsec_numeric_quadratic_modell}).

\subsubsection{$E$/$B$-mode decomposition}

Although we will not apply this formalism we shall now briefly address the so-called $E$/$B$-mode (curl--gradient) decomposition of the intrinsic and 
lensing induced shear field for completeness.
Since the gravitational shear tensor is symmetric and traceless (cf. equation~\ref{eq_ellipticity_tensor} replacing $\epsilon$ by $\gamma$), it proves 
convenient to convert its two degrees of freedom into two \emph{scalar} functions with a definite behaviour under parity transformations
\begin{equation}
 E_\gamma \equiv \nabla^{-2} \left( \nabla^\alpha\nabla^\beta \gamma_{\alpha\beta}\right),
\end{equation}
\begin{equation}
  B_\gamma \equiv  \nabla^{-2} \left( \nabla^\alpha\nabla^\tau \gamma_{\alpha\beta} \varepsilon_{\tau\beta} \right)
\end{equation}
 \citep{1996astro.ph..9149S,1997PhRvD..55.7368K,2002ApJ...568...20C}, where
 \begin{equation}
 \varepsilon_{\tau\beta} \equiv \left(
 \begin{array}{cc}
 0 & \sin\vartheta\\
 -\sin\vartheta & 0
 \end{array}
 \right)
 \end{equation}
and $\nabla^{-2}$ is the inverse Laplace operator on the sphere.
The \emph{gradient} part or $E$-mode is a true scalar, i.e, parity conserving, while the \emph{curl} part or $B$-mode is a pseudo scalar, i.e. changing 
sign under reflections. Any statistical analysis of weak lensing data can then be performed in terms of the angular power spectra of these two fields 
which are invariant under rotations of the reference frame in contrast to the shear components.

The $E$- and $B$-mode of any spin-2 field $\eta$ can be most readily accessed in harmonic space because there is a simple relation between the 
expansion coefficients of the two scalar functions and those of $\eta$ and its complex conjugate $\eta^*$, which carries spin weight $-2$, 
\citep{1999PhRvD..59l3507Z}
\begin{equation}
 E_{\ell m} (k) = -\frac{1}{2}\left( {}_{2}\eta_{\ell m}(k) + {}_{-2}\eta_{\ell m}(k) \right),
\end{equation}
\begin{equation}
 B_{\ell m} (k) = \frac{\mathrm i}{2}\left( {}_{2}\eta_{\ell m}(k) - {}_{-2}\eta_{\ell m}(k) \right).
\end{equation}

Due to the fact that the expansion coefficients of the cosmic shear field and its complex conjugate are identical, i.e. 
${}_2\gamma_{\ell m} = {}_{-2}\gamma_{\ell m}$, the shear field possesses no $B$-mode. Alternatively, this can be seen by 
recalling that any spin-2 quantity $\eta$ may be expressed in terms of a complex potential
\begin{equation}
 \label{eq_spin_two_field_from_complex_potential}
 \eta(\bmath x) = \frac{1}{2}\eth\eth\left(\phi_E (\bmath x) + \mathrm{i} \, \phi_B (\bmath x) \right)
\end{equation}
\citep{1966JMP.....7..863N,2005PhRvD..72b3516C} and realising that the lensing potential is purely real. It turns out that the statistics of the lensing 
$E$-mode is identical to that of the cosmic shear field, i.e. $C^{\gamma\gamma}_\ell(k,k') = C^{E_\gamma E_\gamma}_\ell(k,k')$, provided that the 
expansion coefficients of the potential $\phi_{E_\gamma}$ are appropriately normalised \citep{2005PhRvD..72b3516C}.

The detection of a $B$-mode in weak lensing data may then be used as a diagnostic for systematical errors in the experiment assuming that any other 
potential $B$-mode source can be excluded. Possible $B$-mode contamination results from higher-order lensing effects, like source-lens clustering 
\citep{2002A&A...389..729S}, multiple lensing along the line-of-sight \citep{2010A&A...523A..28K} or violations of the Born-approximation 
\citep{2002ApJ...574...19C,2006JCAP...03..007S}.
Another possible source are intrinsic alignments.
Obviously, the linear model cannot introduce any $B$-mode because in this model the intrinsic ellipticities are obtained from a real potential as well. 
Thus, the same reasoning as in case of cosmic shear applies.
The quadratic model, however, introduces inevitably a non-vanishing $B$-mode as can be seen as follows: 
Looking at equation~\eqref{eq_harmonic_coeff_quadratic_model} and realising that
\begin{equation}
\eth\eth Y_{\ell m} (\n) = \sqrt{\frac{(\ell+2)!}{(\ell-2)!}} {}_{2}Y_{\ell m} (\n),
\end{equation}
\begin{equation}
 \bar\eth\bar\eth Y_{\ell m} (\n) = \sqrt{\frac{(\ell+2)!}{(\ell-2)!}} {}_{-2}Y_{\ell m} (\n)
\end{equation}
we note that the only potentially spin-dependent part is given by the angular mixing matrix defined in equation~\eqref{eq_angular_mixing_matrix}. Carrying out the decomposition of $\epsilon^*(\bmath x)=\epsilon_+(\bmath x) - \mathrm i \epsilon_\times (\bmath x)$ one readily finds
\begin{equation}
 {}_{-2}\mathcal{W}_{m_1 m_2 m}^{\ell_1 \ell_2 \ell} = (-1)^{\ell_1+\ell_2+\ell}  {}_{2}\mathcal{W}_{m_1 m_2 m}^{\ell_1 \ell_2 \ell}
\end{equation}
and therefore ${}_{2}\epsilon_{\ell m}(k) \neq {}_{-2}\epsilon_{\ell m}(k)$.
Consequently, for the quadratic model the $B$-mode of the ellipticity field does not vanish.

Finally, we would like to mention that there are physical processes which transfer power from the $E$-mode of the intrinsic ellipticity field into its $B$-
mode and vice versa. Recently, \citet{2013MNRAS.428.1312G,2013arXiv1302.2607G} investigated this $E/B$-mode conversion due to peculiar motion 
and gravitational lensing, respectively, in the flat-sky limit.

\section{Numerical results}
\label{sec_numerical_results}

\subsection{Linear model}
\label{subsec_numeric_linear_model}

After the derivation of the various covariance matrices in the last section we would now like to illustrate our findings by presenting some numerical 
results.
We focus on the linear alignment model, which in contrast to the quadratic model, exhibits both \textit{II}- and \textit{GI}-alignments.
In general, the $3d$ weak shear formalism is numerically very demanding due to the large number of $k$-integrations required. A further complication 
arises from the fact that the functions to be integrated contain spherical Bessel functions which are highly oscillatory. \citet{2012MNRAS.422.3056A} 
showed how the numerical efforts can be minimised. We adapt their strategy in our numerical evaluation of the covariance matrices of the \textit{II}- and 
\textit{GI}-alignments predicted by the linear model.

In order to make the computation feasible we restrict our analysis to a (discrete) $k$-space interval from 
$k_{\mathrm{min}}=10^{-3} \ \mathrm{Mpc}^{-1}$ to $k_{\mathrm{max}}=10^{-1} \ \mathrm{Mpc}^{-1}$ covered by $N_k=200$ equidistant steps. We 
therefore do not apply any additional filtering to the CDM power spectrum. Likewise, the finite wavevector interval corresponds to a sharp cutoff filter in 
Fourier space which has been used by \citet{2004PhRvD..70f3526H} in their two-dimensional analysis. Furthermore, we consider $N_z=1000$ steps in 
redshift space ranging from $z_{\mathrm{mim}} = 10^{-4}$ to $z_{\mathrm{max}}=10$.

In accordance with \citet{2004PhRvD..70f3526H,2007NJPh....9..444B,2011A&A...527A..26J} we choose the value for the constant of the linear 
alignment model, $C_1$, in such a way that it matches the SuperCOSMOS observations at low redshift \citep{2002MNRAS.333..501B}: 
$C_1 H_0^2 = 8.93 \times 10^{-3}$ (we use natural units where the speed of light is set to unity).

\begin{figure}
 \label{fig_diagonal_elements_cov_matrices_z_0_0_21_l_10_20}
 \centering
 \resizebox{\hsize}{!}{\includegraphics[]{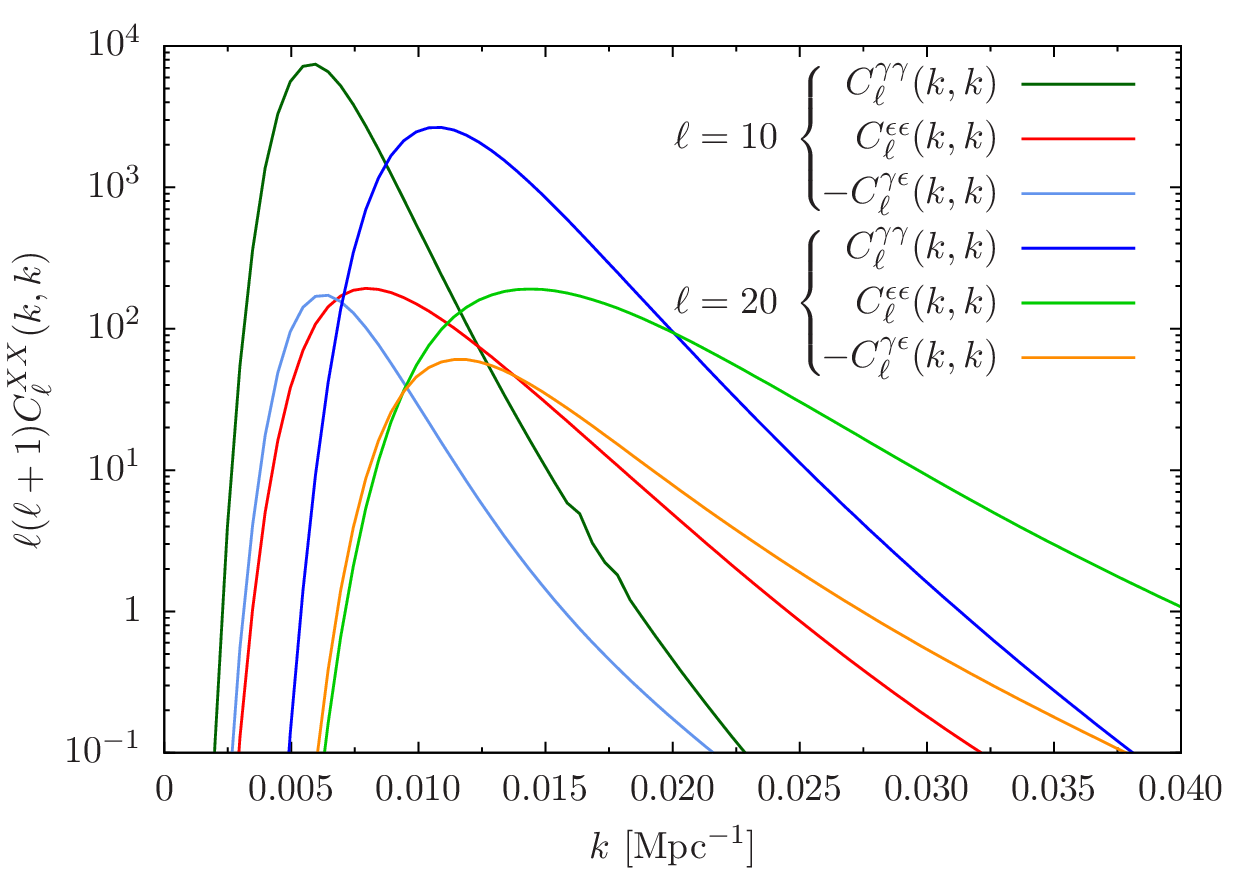}}
 \caption{Diagonal elements of the covariance matrices of the cosmic shear field, \textit{II}-  and \textit{GI}-alignments for the linear model. We show the 
 results for multipoles $\ell=10$ and $\ell=20$. The median redshift has been set to $z_{\mathrm{median}} = 0.3$.}
\end{figure}

\begin{figure}
 \label{fig_diagonal_elements_cov_matrices_z_0_0_21_l_100_200}
 \centering
 \resizebox{\hsize}{!}{\includegraphics[]{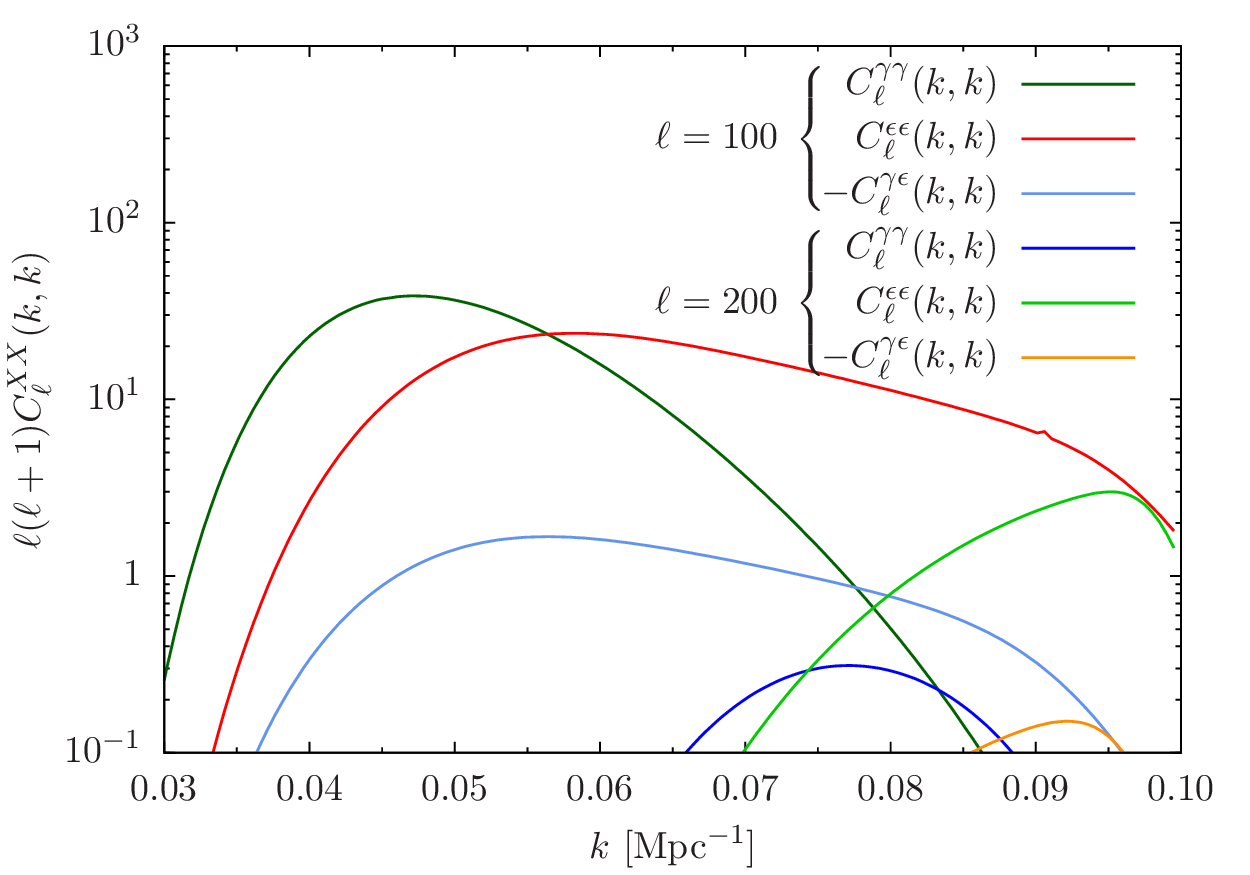}}
 \caption{Diagonal elements of the covariance matrices of the cosmic shear field, \textit{II}- and \textit{GI}-alignments for the linear model. We show the results for multipoles $\ell=100$ and $\ell=200$. The median redshift has been set to $z_{\mathrm{median}} = 0.3$.}
\end{figure}

First, we concentrate on the diagonal elements of the various covariance matrices.
In Figure~\ref{fig_diagonal_elements_cov_matrices_z_0_0_21_l_10_20} and~\ref{fig_diagonal_elements_cov_matrices_z_0_0_21_l_100_200} we 
show the results for four different multipoles, $\ell=10$, 20, 100 and 200,  assuming a shallow survey with $z_{\mathrm{median}} = 0.3$.
Note the different $k$-range in each of the plots. We only depict the range where the amplitude of the covariance matrix is appreciable.
The first observation is the expected trend that the larger the multipole order the larger the $k$-value where the covariance matrix peaks.
Interestingly, the peak position slightly differs for all three covariance matrices. The differences in the peak location increases with increasing multipole 
order.
At low multipoles ($\ell = 10$, 20 corresponding to an angular scale of about ten degrees) the weak lensing signal  dominates. It exceeds the signal of 
both intrinsic alignment types by more than one order of magnitude. For the smallest multipole the amplitude of the \textit{II}-alignments is compatible to 
that of the \textit{GI}-alignments but then starts to dominate the alignment signal and finally exceeds it by more than one order of magnitude for 
$\ell = 200$. On this angular scale the \textit{GI}-contribution is almost as large as the cosmic shear signal. Consequently, for $\ell=200$ the total 
covariance matrix is dominated by the \textit{II}-alignment term.

\begin{figure}
 \label{fig_diagonal_elements_cov_matrices_z_0_0_64_l_10_20}
 \centering
 \resizebox{\hsize}{!}{\includegraphics[]{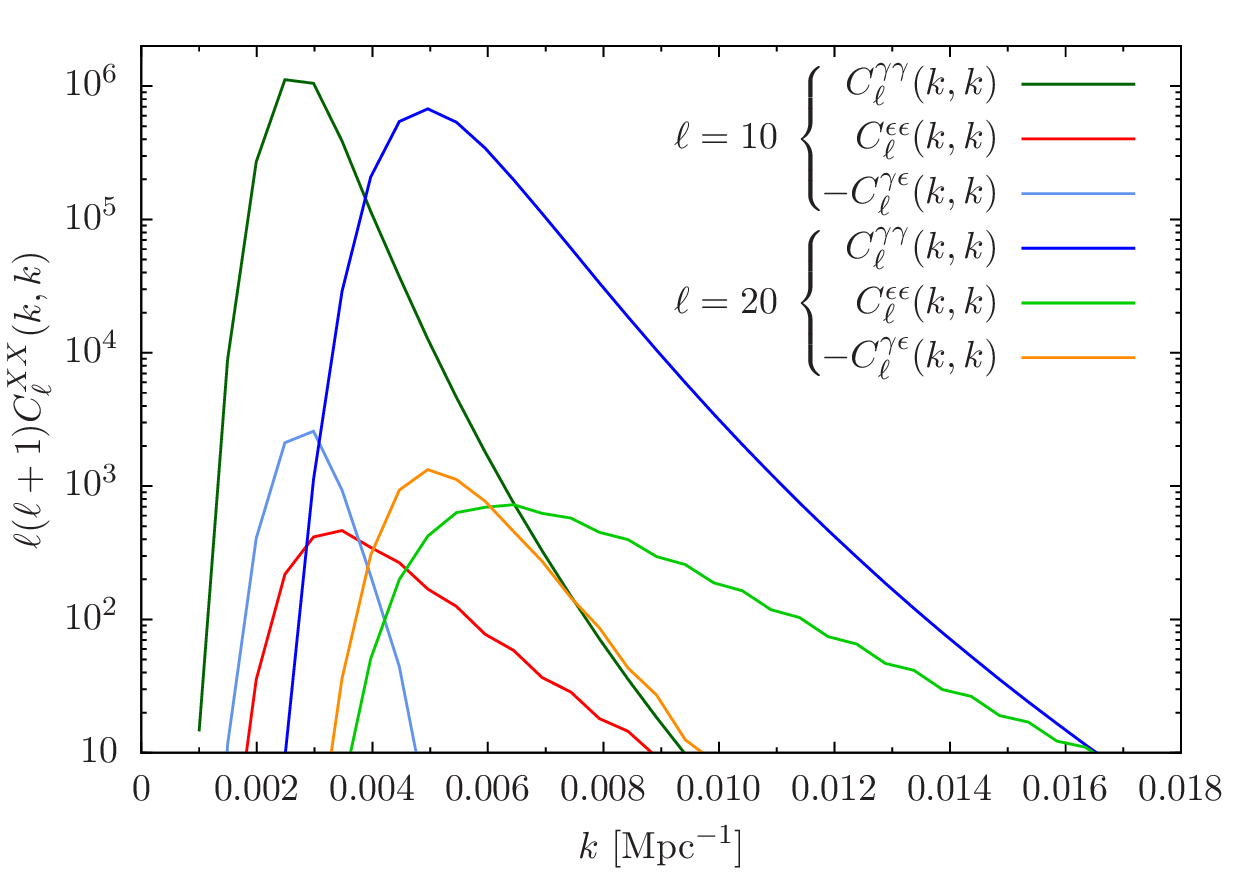}}
 \caption{Diagonal elements of the covariance matrices of the cosmic shear field, \textit{II}-  and \textit{GI}-alignments for the linear model. We show the 
 results for the multipoles $\ell=10$ and $\ell=20$. In contrast to the shallow survey assumed in 
 Figure~\ref{fig_diagonal_elements_cov_matrices_z_0_0_21_l_10_20} the median redshift has now been set to $z_{\mathrm{median}} = 0.9$ as 
 anticipated by the \textit{Euclid} mission.}
\end{figure}

\begin{figure}
 \label{fig_diagonal_elements_cov_matrices_z_0_0_64_l_100_200}
 \centering
 \resizebox{\hsize}{!}{\includegraphics[]{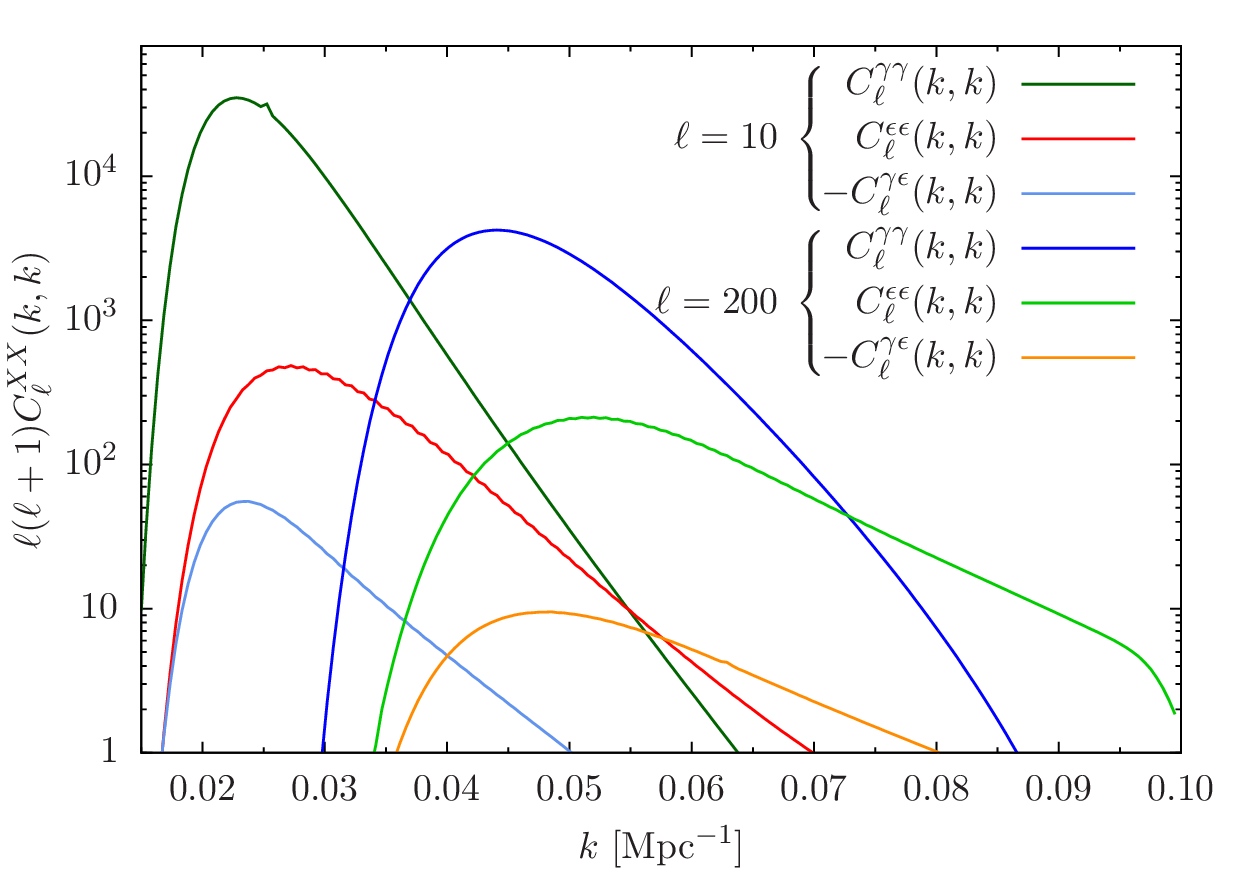}}
 \caption{Diagonal elements of the covariance matrices of the cosmic shear field, \textit{II}- and \textit{GI}-alignments for the linear model. We show the 
 results for the multipoles $\ell=100$ and $\ell=200$. In contrast to the shallow survey assumed in 
 Figure~\ref{fig_diagonal_elements_cov_matrices_z_0_0_21_l_100_200} the median redshift has now been set to $z_{\mathrm{median}} = 0.9$ as 
 anticipated by the \textit{Euclid} mission.}
\end{figure}

The situation is different for a deeper survey. In Figure~\ref{fig_diagonal_elements_cov_matrices_z_0_0_64_l_10_20} 
and~\ref{fig_diagonal_elements_cov_matrices_z_0_0_64_l_100_200} we show the results for a survey with $z_{\mathrm{median}} = 0.9$ having an 
eye on the \textit{Euclid} mission. 
As before, we plot the diagonal elements of all three covariance matrices for $\ell=10$, 20, 100 and 200. 
Again only the $k$-range featuring substantial amplitude is shown for clarity.
We find that weak lensing contributes most to the diagonal elements of the covariance matrices for all four multipoles under consideration. The 
difference with respect to the \textit{II}-contributions decreases with multipole order (from roughly three orders of magnitude for 
$\ell=10$ to one order of magnitude for $\ell =200$). While on large angular scales the \textit{GI}-contributions are slightly enhanced in comparison to 
those from the \textit{II}-alignments, the latter dominate vastly on smaller scales where the \textit{II}-contributions exceed the \textit{GI}-signal by more 
than one order of magnitude. Furthermore, we see that the differences in the peak positions are much lees pronounced than for the shallow survey.

Confronting the results obtained for the two different surveys one immediately recognises that the maxima are shifted to larger scales in case of the 
higher redshift sample. 
It is interesting to note that for both survey specifications \textit{II}-alignments on the one hand and weak cosmic shear and \textit{GI}-alignments on the 
other hand exhibit a different dependence on the wave number $k$. The weak lensing and \textit{GI}-signal fall off much steeper than that of the 
\textit{II}-alignments.

A direct comparison of our results to the two-dimensional analysis carried out by \citet{2004PhRvD..70f3526H} is of course not possible. 
However, we can draw some general, qualitative conclusions. As \citet{2004PhRvD..70f3526H}, we find that in case of a shallow survey the \textit{II}-
contributions dominate over the weak cosmic shear signal as well as the \textit{GI}-signal. Increasing the depth of the survey the situation is reversed 
and lensing is the dominant contribution. This is also true for the two-dimensional power spectrum. However, as an important difference with respect to 
the work of \citet{2004PhRvD..70f3526H}
we find that for the deeper survey the major contamination of the weak lensing signal is not given by the \textit{GI}-alignments but by the \textit{II}-
alignments.

\begin{figure*}
 \label{fig_covariance_matrices_linear_model}
  \centering%
\begin{minipage}{0.04\textwidth}
 	\raggedleft\rotatebox{90}{\large cosmic shear}
 \end{minipage}%
\begin{minipage}{0.48\textwidth}
{\centering%
 \large$\ell = 10$\\}
 \vspace{0.5\baselineskip}
 \raggedright
 \qquad\includegraphics[scale=0.825]{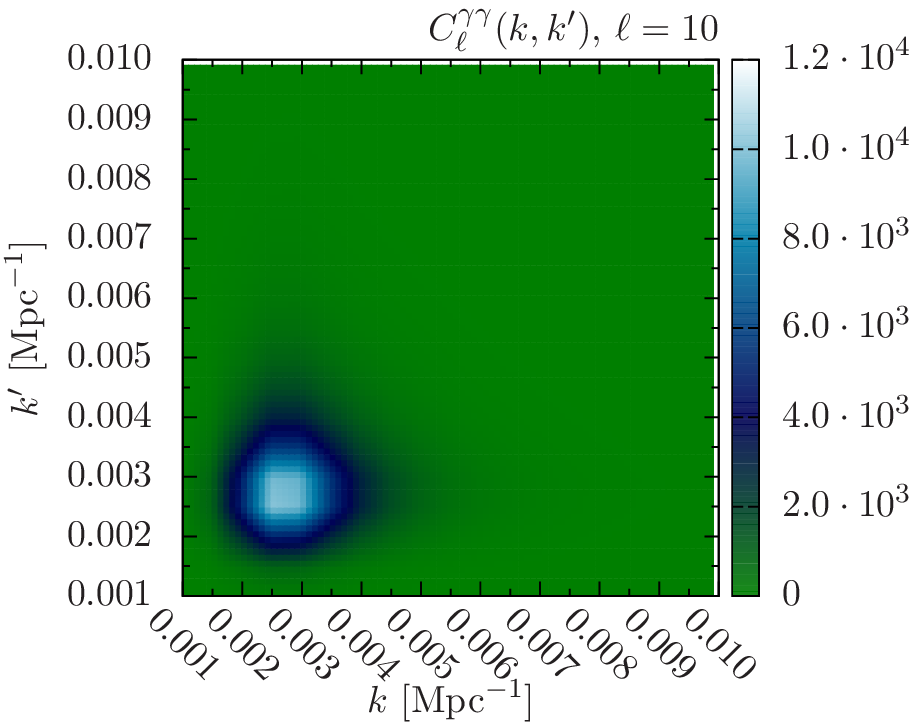}
\end{minipage}%
 \begin{minipage}{0.48\textwidth}
 {\centering%
 \large$\ell = 100$\\}
 \vspace{0.5\baselineskip}
 \raggedright
 \qquad\includegraphics[scale=0.825]{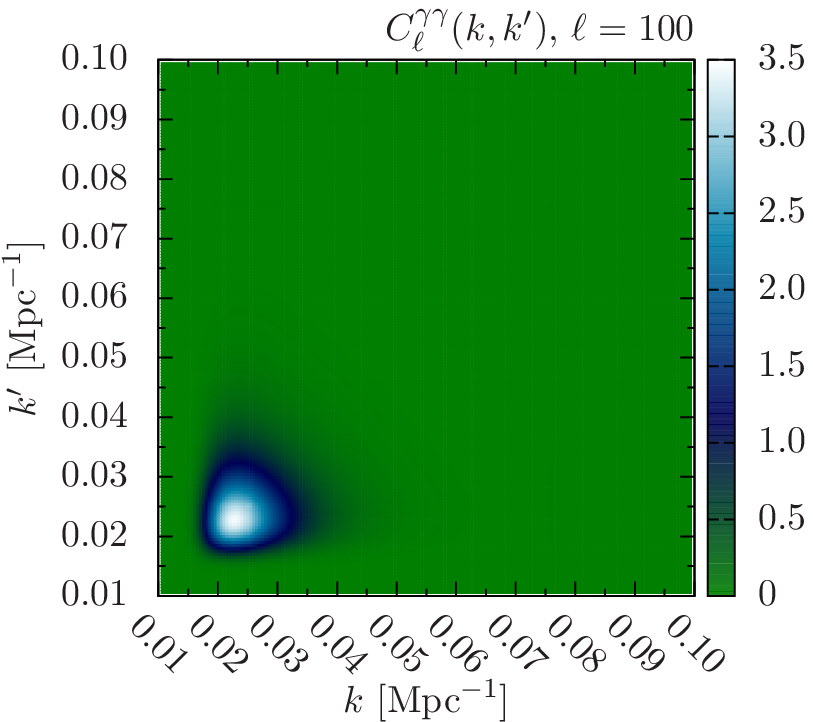}
 \end{minipage}%
 \\
 \begin{minipage}{0.04\textwidth}
 	\raggedleft\rotatebox{90}{\large\textit{II}-alignments}
 \end{minipage}%
\begin{minipage}{0.48\textwidth}
 \raggedright%
 \vspace{0.5\baselineskip}
 \qquad\includegraphics[scale=0.825]{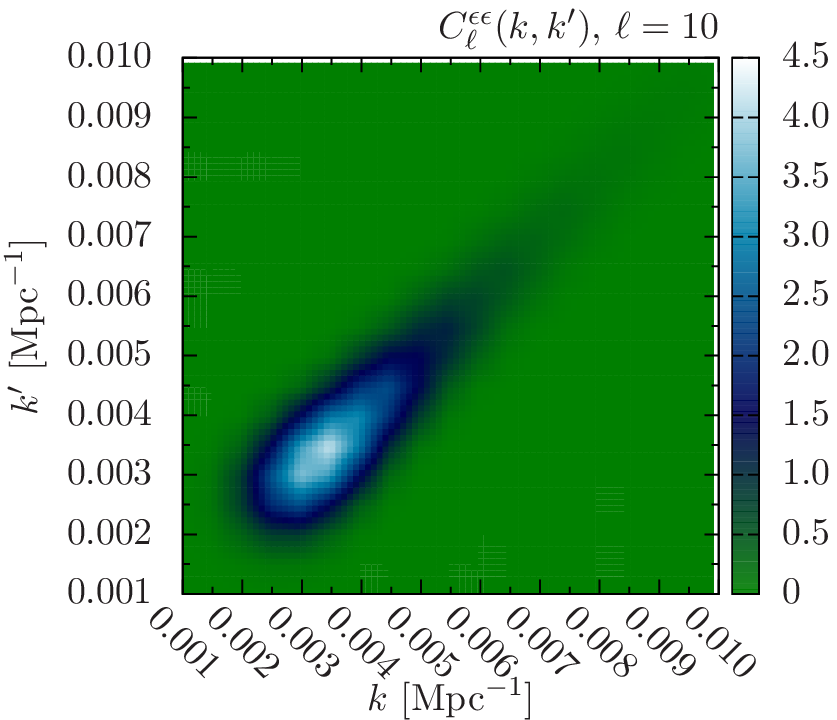}
\end{minipage}%
 \begin{minipage}{0.48\textwidth}
 \raggedright%
  \vspace{0.5\baselineskip}
 \qquad\includegraphics[scale=0.825]{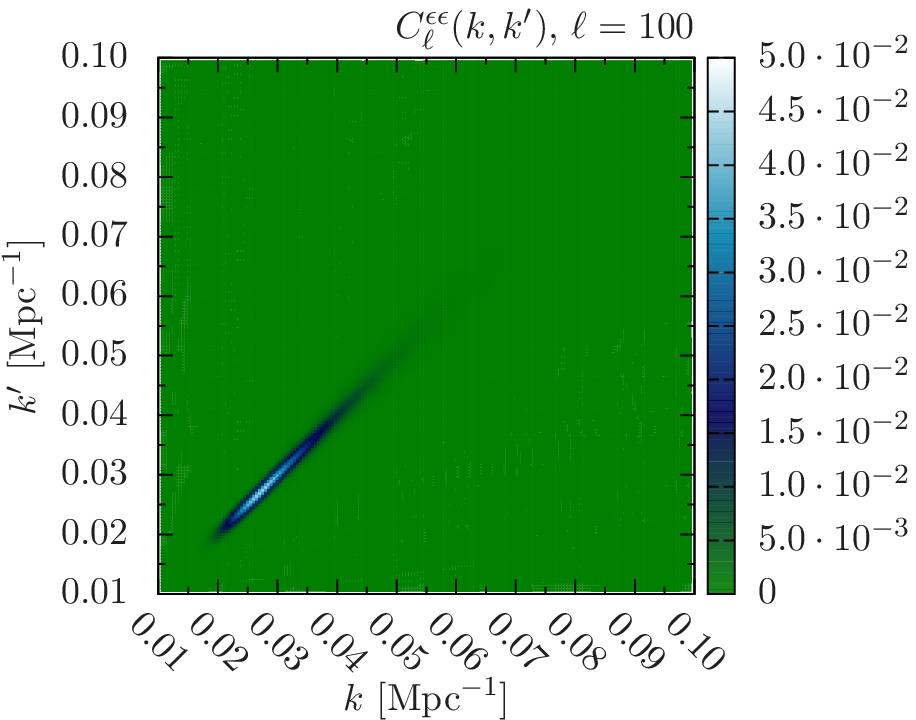}
 \end{minipage}%
 \\
 \begin{minipage}{0.04\textwidth}
 	\raggedleft\rotatebox{90}{\large\textit{GI}-alignments}
 \end{minipage}%
\begin{minipage}{0.48\textwidth}
 \raggedright%
 \vspace{0.5\baselineskip}
 \qquad\includegraphics[scale=0.825]{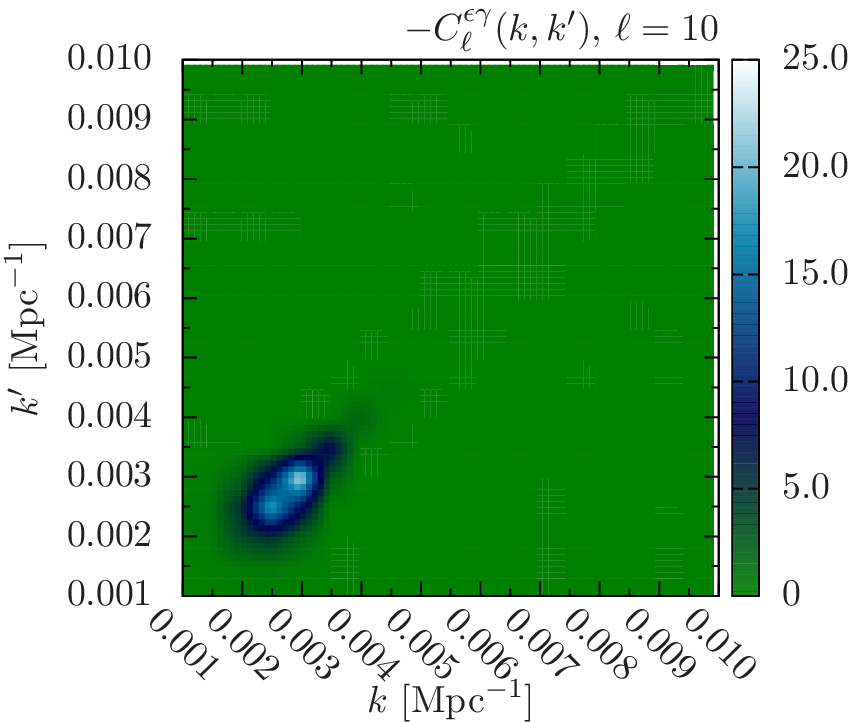}
\end{minipage}%
 \begin{minipage}{0.48\textwidth}
 \raggedright%
  \vspace{0.5\baselineskip}
 \qquad\includegraphics[scale=0.825]{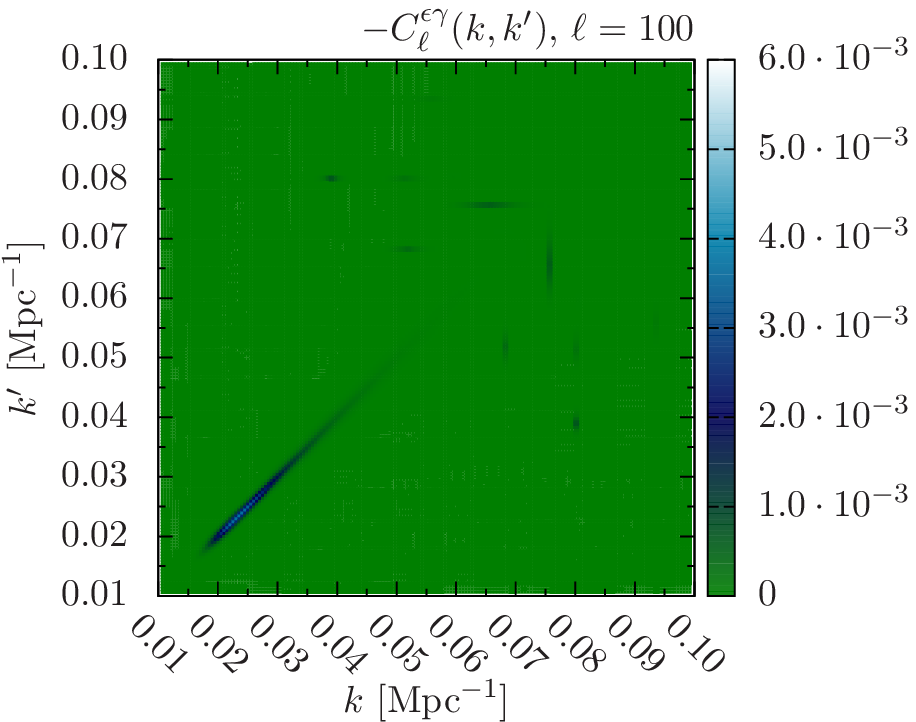}
 \end{minipage}%
 \caption{Covariance matrices for the linear alignment model. From top to bottom: covariance of the lensing induced shear field (first row), \textit{II}-
 alignments (second row) and \textit{GI}-alignments (third row) for multipoles $\ell=10$ (left column) and $\ell=100$ (right column). The median redshift 
 has been chosen to be $z_{\mathrm{median}} = 0.9$.}
\end{figure*}

Finally, we look at the shape of the full three-dimensional covariance matrices. In Figure~\ref{fig_covariance_matrices_linear_model} we show from top 
to bottom the covariance matrices of the cosmic shear field, \textit{II}-alignments and \textit{GI}-alignments for multipoles $\ell=10$ (left panel) and 
$\ell=100$ (right panel), respectively. Here we only consider the \textit{Euclid}-like high redshift sample.
Note we do not show the entire $k$-range but concentrate on the part from which the covariance matrices receive appreciable contributions.
As already observed before, it is common to all three covariance matrices that the higher the multipole order the more the signal is shifted towards 
smaller scales. The individual shape of the matrices, however, is quite different. While the weak lensing signal is rather circular, the covariance of the 
\textit{II}-alignments are elongated along the diagonal of the $k$-plane. This corresponds to the slow decline of the \textit{II}-signal with increasing 
wavenumber we have already noticed during our discussion of the diagonal entries above. In contrast to this, the \textit{GI}-contributions are much more 
compact. Their extent in the $k$-plane is several times smaller than that of the \textit{II}-alignments. Interestingly, for $\ell = 100$ the contributions of 
both alignment types are almost completely concentrated on the diagonal, whereas the shape of the cosmic shear covariance resembles that of an 
isosceles triangle. Especially the \textit{GI}-alignments become extremely narrow in the $k$-plane.

These findings allow a first qualitative estimate of how intrinsic alignments may bias cosmological parameters inferred from $3d$ cosmic shear studies. 
With respect to the pure cosmic shear signal the total covariance matrix, i.e. the sum of all three matrices, would be tilted towards larger $k$-modes 
since the contributions from the \textit{II}-alignments add power on smaller scales, whereas those of the \textit{GI}-alignments reduce the amplitude at 
smaller wave numbers. This tilt suggests that in particular the normalisation of the power spectrum~$\sigma_8$, the matter 
fraction~$\Omega_\mathrm{m}$ and the Hubble parameter~$h$ are biased. The actual magnitude of this bias, however, is hard to predict because the 
effect of power enhancement and suppression is different for different multipoles.

\subsection{Quadratic model}
\label{subsec_numeric_quadratic_modell}

There are two main complications in evaluating the quadratic model numerically, the double sum in 
equation~\eqref{eq_covariance_matrix_quadratic_model} and the mode coupling integral $\tilde{I}_{\ell_1 \ell_2\ell}(k_1,k_2,k)$ defined by 
equation~\eqref{eq_mode_coupling_I_tilde}. The double sum results from the product of two gradient fields 
(cf. equation~\ref{eq_epsilon_quadratic_model_eth}); the corresponding multipoles have to be considered as a kind of inner indices. Therefore, a 
truncation of the sum at relatively low multipoles, $\ell_{1,2}\sim \mathcal{O}(10^2)$, is not permissible. The very fact that this sum converges rather 
slowly leaves the numerical implementation at least challenging.

The situation is further aggravated by the evaluation of the mode coupling integral $\tilde{I}_{\ell_1 \ell_2 \ell} (k_1, k_2, k)$. For large multipoles, as 
required by the double sum in order to converge, the computation is numerically problematic due to the oscillatory behaviour of higher-order spherical 
Bessel functions.
This kind of integrals appears in nuclear scattering theory and analytical attempts to its evaluation exist 
\citep{1991JPhA...24.1435M,2010JPhA...43S5204M}. These analytical expressions, however, contain multiple sums over Wigner $3j$- 
and $6j$-symbols as well as associated Legendre functions making the practical use for an efficient computation of a large number of these integrals 
quite limited.
A more promising approach might be to resort to the asymptotic forms of spherical Bessel functions in the high $\ell$-limit. Though this obviously 
requires an elaborate distinction of cases since the multipoles enter the sum of equation~\eqref{eq_covariance_matrix_quadratic_model} in all possible 
combinations.
Aiming at the constitutive formalism for a description of intrinsic alignments in the framework of $3d$ cosmic shear the explicit development of such a 
scheme is beyond the scope of this paper and is reserved for future work.

\section{Summary}
\label{sec_summary}

We developed a formalism to incorporate the intrinsic alignment of galaxy shapes into the framework of $3d$ cosmic shear.
For the description of the intrinsic galaxy ellipticities we used two different models being linear and quadratic in the cosmic tidal field, respectively.
\begin{enumerate}
 \item Introducing an appropriate auxiliary scalar potential, we were able to formulate the intrinsic ellipticity field for each of the two models in terms of 
 second eth-derivatives emphasising the spin-2 character of the ellipticity field.
 \item We then derived expressions for the covariance matrices of the intrinsic alignments. Both ellipticity models give rise to \textit{II}-alignments but for 
 Gaussian density fluctuations \textit{GI}-alignments are only present in the linear model.
 \item In contrast to conventional  two-dimensional weak lensing analyses the covariance matrix of the three-dimensional shear field acquires 
 off-diagonal elements in $k$-space reflecting the fact that the lensing potential is not a statistically homogeneous field. The same is true for the 
 covariance matrices of the \textit{II}- and \textit{GI}-alignments. Statistical homogeneity, however, is maintained in the angular parts of the ellipticity and 
 shear field, respectively, in case of an idealised observation with access to the full sky. As a consequence, the corresponding covariance matrices are 
 diagonal with respect to the angular multipoles $\ell$ and $m$.
 \item In addition, we considered the $E/B$-mode decomposition of the ellipticity field and found that in case of the linear model the $B$-mode is 
 identically zero. Thus, for the linear model the intrinsic ellipticities exhibit the same behaviour under parity transformations as those induced by lensing. 
 Opposed to that, the quadratic model does possess a non-vanishing parity odd $B$-mode.
 \item We evaluated numerically the covariance matrices of both \textit{II}- and \textit{GI}-alignments for the linear model and compared it to that of 
 cosmic shear. In case of a shallow survey ($z_{\mathrm{median}} = 0.3$) we found that for large multipoles the \textit{II}-alignments dominate the 
 signal underlining the fact that intrinsic alignments are a small scale phenomenon.
 For a deep redshift survey like \textit{Euclid} ($z_{\mathrm{median}} = 0.9$), however, the lensing signal becomes dominant for all multipoles. In this 
 case, the \textit{II}-alignment contamination is more than one order of magnitude smaller with respect to the shear signal. Furthermore, we showed that 
 in contrast to two-dimensional lensing studies the \textit{GI}-contributions are only slightly enhanced if at all in comparison to the \textit{II}-alignments 
 which are the dominant contaminant on almost all scales for both survey specifications.
 \item Finally, we investigated the shape of the various covariance matrices in the $k$-$k'$-plane. Concentrating on a \textit{Euclid}-like survey we found 
 a fundamental difference between the covariance matrices of cosmic shear on the one hand and of \textit{II}- and \textit{GI}-alignments, respectively, on 
 the other hand. While the shape of the lensing covariance matrix is rather circular that of the intrinsic alignments is strongly elongated along the 
 diagonal $k=k'$. For large multipoles ($\ell\sim100$) the covariance matrices of both alignment types tend to almost completely concentrate along this 
 diagonal.
\end{enumerate}

Having established the basic formalism for a consistent treatment of the linear and quadratic alignment model in the framework of $3d$ weak lensing, 
future efforts should now be addressed to a detailed quantitative study of their impact on cosmological parameters derived from $3d$ cosmic shear 
measurements.

\section*{Acknowledgements}

We would like to thank Youness Ayaita and Maik Weber for providing us with a version of their excellent $3d$ weak lensing code.
PhMM acknowledges funding from the Graduate Academy Heidelberg and support from the International Max Planck
Research School for Astronomy and Cosmic Physics in Heidelberg as well as from the Heidelberg Graduate School of Fundamental Physics.
BMS's work is supported by the German Research Foundation (DFG) within the framework of the excellence initiative through the Heidelberg Graduate 
School of Fundamental Physics. We are in particular thankful to the anonymous referee for thoughtful comments which helped to improve our 
presentation.

\bibliography{bibtex/aamnem,bibtex/references}
\bibliographystyle{mn2e}

\appendix
\section{Derivation of the covariance matrix of the quadratic model}
\label{sec_appendix}
In this appendix we fill up the missing steps in the derivation of equation~\eqref{eq_covariance_matrix_quadratic_model}. In order to simplify our notation we denote multipoles with a bar $\bar\ell$ when we refer to $\ell$ and $m$ at the same time. We begin with the computation of $\langle \epsilon_{\ell m}(k) \epsilon^*_{\ell' m'}(k')\rangle$.
Denoting the trispectrum of the Newtonian gravitational potential evaluated today by
\begin{equation}
 \mathbb T^{\bar\ell_1\bar\ell_2}_{\bar\ell_1'\bar\ell_2'}(k_1, k_2, k_1', k_2')\equiv
 \bigl\langle \Phi^{{\mathcal S},0}_{\ell_1 m_1} (k_1) \Phi^{{\mathcal S},0}_{\ell_2 m_2} (k_2) \Phi^{{\mathcal S},0*}_{\ell'_1 m'_1} (k_1') \Phi^{{\mathcal S},0*}_{\ell'_2 m'_2} (k'_2)\bigr\rangle
\end{equation}
we have
\begin{eqnarray}
 \nonumber
  \left\langle \epsilon_{\ell m}(k) \epsilon^*_{\ell' m'}(k')\right\rangle &=& 
 C_2^2 \sum_{\bar\ell_1\bar\ell_2\bar\ell'_1\bar\ell'_2}
 \bar{\mathcal{Q}}^k_{\bar\ell_1\bar\ell_2\bar\ell}(k_1,k_2) \bar{\mathcal{Q}}^{k'}_{\bar\ell'_1\bar\ell'_2\bar\ell'}(k'_1,k'_2)\\
 &&
 \times\,\mathbb T^{\bar{\ell}_1\bar\ell_2}_{\bar\ell_1'\bar\ell_2'}(k_1, k_2, k_1', k_2').
\end{eqnarray}
In the last equation we have defined
\begin{eqnarray}
 \nonumber
  \bar{\mathcal{Q}}^k_{\bar\ell_1\bar\ell_2\bar\ell}(k_1,k_2) &\equiv&
 (-1)^{m+1} \sqrt{\frac{2}{\upi^3}} \ell_1(\ell_1+1) \sqrt{\frac{(\ell_2+2)!}{(\ell_2-2)!}} \mathcal{W}^{\ell_1\ell_2\ell}_{m_1 m_2 m} \\
 &&\label{eq_Q_bar}
 \times\,\mathcal{I}_{\ell_1\ell_2\ell} (k_3,k_4,k)\, \tilde\eta_{\ell_1} (k_3,k_1)\, \tilde\eta_{\ell_2} (k_4,k_2).
\end{eqnarray}
Exploiting the orthogonality of the spherical Bessel functions
\begin{equation}
 \label{eq_orthogonality_relation_Bessel_functions}
 \int k^2\dd k \, j_\ell (k\chi) j_\ell (k\chi') = \frac{\upi}{2\chi^2}\delta_\mathrm{D}(\chi - \chi')
\end{equation}
\citep{1972hmfw.book.....A} the matrix product in equation~\eqref{eq_Q_bar} can be simplified considerably. Introducing
\begin{equation}
 \label{eq_mode_coupling_I_tilde}
 \tilde{\mathcal{I}}_{\ell_1\ell_2\ell}(k_1,k_2,k) \equiv \frac{\upi}{2}\int\frac{\dd\chi}{\chi^2} j_{\ell_1}(k_1\chi) j_{\ell_2}(k_2\chi)j_{\ell}(k\chi)
\end{equation}
we are left with
\begin{eqnarray}
 \nonumber
  \bar{\mathcal{Q}}^k_{\bar\ell_1\bar\ell_2\bar\ell}(k_1,k_2) &\equiv&
 (-1)^{m+1} \sqrt{\frac{2}{\upi^3}} \ell_1(\ell_1+1) \sqrt{\frac{(\ell_2+2)!}{(\ell_2-2)!}} \mathcal{W}^{\ell_1\ell_2\ell}_{m_1 m_2 m} \\
 &&
 \times\,\tilde{\mathcal{I}}_{\ell_1\ell_2\ell} (k_1,k_2,k).
\end{eqnarray}
As explained in Section~\ref{subsec_quadratic_model} the trispectrum is evaluated via Wick's theorem.
The trispectrum is then diagonal in the respective combinations of $k_i$ and $\bar\ell_i$. Therefore, the sums over $\bar\ell'_{1,2}$ collapse. The remaining sums over $m_{1,2}$ trivialises due to the fact that
\begin{equation}
 \sum_{m_1m_2}
 \left(
 \begin{array}{ccc}
  \ell_1 & \ell_2 & \ell\\
  m_1 & m_2 & m
 \end{array}
 \right)
  \left(
 \begin{array}{ccc}
  \ell_1 & \ell_2 & \ell'\\
  m_1 & m_2 & m'
 \end{array}
 \right)
 =
 \frac{1}{2\ell +1} \delta_{\ell\ell'}\delta_{mm'}.
\end{equation}
Having removed the entire $m_{1,2}$-dependence we now set
\begin{eqnarray}
 \nonumber
 \mathcal{Q}^k_{\ell_1\ell_2\ell}(k_1,k_2) &\equiv&
 \sqrt{\frac{2}{\upi^3}} \ell_1(\ell_1+1) \sqrt{\frac{(\ell_2+2)!}{(\ell_2-2)!}}
 \left(
 \begin{array}{ccc}
  \ell_1 & \ell_2 & \ell\\
 0 & -2 & 2
 \end{array}
 \right)\\
 &&
 \times \, \Pi_{\ell_1\ell_2\ell}\tilde{\mathcal{I}}_{\ell_1\ell_2\ell} (k_1,k_2,k)
\end{eqnarray}
and exploit another property of the Wigner $3j$-symbols, namely
\begin{equation}
 \left(
 \begin{array}{ccc}
  \ell_1 & \ell_2 & \ell\\
  m_1 & m_2 & m
 \end{array}
 \right)
 =(-1)^{\ell_1 + \ell_2 + \ell}
  \left(
 \begin{array}{ccc}
  \ell_2 & \ell_1 & \ell\\
  m_2 & m_1 & m
 \end{array}
 \right),
\end{equation}
to arrive at
\begin{eqnarray}
 \nonumber
 \left\langle \epsilon_{\ell m}(k) \epsilon^*_{\ell' m'}(k')\right\rangle &=&
 \frac{A^4C_2^2}{2\ell + 1} \sum_{\ell_1\ell_2} \mathcal{Q}^k_{\ell_1\ell_2\ell} (k_3,k_4) \frac{P(k_3)}{k_3^4} \frac{P(k_4)}{k_4^4}\\
 &&
 \nonumber
 \times \left(\mathcal{Q}^{k'}_{\ell_1\ell_2\ell} (k_3,k_4 ) + (-1)^{\ell_1+\ell_2+\ell} \, \mathcal{Q}^{k'}_{\ell_2\ell_1\ell}(k_3,k_4) \right).\\
 && 
\end{eqnarray}
Equation~\eqref{eq_covariance_matrix_quadratic_model} may then be recovered by setting
\begin{equation}
 \label{eq_def_B_matrix_quadratic_model}
 \mathcal B^k_{\ell\ell_1\ell_2}(k_3,k_4) \equiv \mathcal Z_\ell ( k, k'') \mathcal M_\ell(k'',k''') \mathcal Q^{k'''}_{\ell_1\ell_2\ell}(k_3,k_4).
\end{equation}

\bsp

\label{lastpage}

\end{document}